\documentclass[fleqn,usenatbib]{mnras}
\pdfoutput=1 

\usepackage[T1]{fontenc}
\usepackage{ae,aecompl}

\usepackage{graphicx}	
\usepackage{amsmath}	
\usepackage{amssymb}	
\usepackage{lscape}
\usepackage[flushleft]{threeparttable}
\usepackage[usenames]{color}

\newcommand{\sm}{\sim\!}
\newcommand{\kms}{\mbox{km\ s}$^{-1}$}
\newcommand{\lsun}{L_\odot}
\newcommand{\msun}{M_\odot}

\newcommand{\msu}{M^{\rm su}}
\newcommand{\rsu}{R^{\rm su}}
\newcommand{\vsu}{V^{\rm su}}
\newcommand{\tsu}{T^{\rm su}}
\newcommand{\mc}{M^*}
\newcommand{\lc}{L^*}
\newcommand{\lcIR}{L^{*}_{\rm NIR}}
\newcommand{\magn}{{\cal M}}
\newcommand{\ms}{M_\star}

\newcommand{\msg}{M_{\star, \rm{gr}}}
\newcommand{\Mtwel}{M_{12}}

\newcommand{\Rd}{R_{\rm d}}

\newcommand{\rs}{r_{\rm s}}
\newcommand{\Ms}{M_{\rm s}}
\newcommand{\rH}{r_{\rm H}}

\newcommand{\Rvir}{R_{\rm vir}}

\newcommand{\Vvir}{V_{\rm vir}}
\newcommand{\Mvir}{M_{\rm vir}}

\newcommand{\Dvir}{\Delta_{\rm vir}}

\newcommand{\Mp}{M_{\rm p}}
\newcommand{\Reff}{R_{\rm e}}
\newcommand{\sige}{\sigma_{\rm e}}
\newcommand{\SBe}{\mu_{\rm e}}
\newcommand{\ti}{t_{\rm i}}
\newcommand{\tf}{t_{\rm f}}

\newcommand{\hi}{{H$\,$\footnotesize I}}

\newcommand{\deli}{\delta_{\rm i}}

\newcommand{\rhou}{\bar{\rho}_{\rm{univ}}}
\newcommand{\rhoc}{\rho_{\rm{crit}}}
\newcommand{\zcoll}{z_{\rm{coll}}}
\newcommand{\zi}{z_{\rm{i}}}
\newcommand{\zf}{z_{\rm{f}}}
\newcommand{\zo}{z_{\rm{obs}}}
\newcommand{\Om}{\Omega_{\rm m}}

\newcommand{\Ol}{\Omega_{\rm \Lambda}}
\newcommand{\Omo}{\Omega_{{\rm m},0}}

\newcommand{\Olo}{\Omega_{\Lambda,0}}
\newcommand{\Op}{\Omega_{\rm p}}

\newcommand{\Ngal}{N_{\rm gal}}

\newcommand{\magal}{\textsc{MaGaLie}}
\newcommand{\nemo}{\textsc{NEMO}}
\newcommand{\bdgal}{\textsc{BUILDGAL}}

\title[First-ranked ellipticals from hierarchical dry merging]{Forming
  first-ranked early-type galaxies through hierarchical
  dissipationless merging}

\author[J.~M.\ Solanes et al.]{\parbox{\textwidth}{
Jos\'e M.\ Solanes,$^{1}$\thanks{E-mail: \texttt{jm.solanes@ub.edu}}
Jaime D.\ Perea,$^{2}$
Laura Darriba,$^{1}$
Carlos Garc\'\i a-G\'omez,$^{3}$
Albert Bosma,$^{4}$
and Evangelia Athanassoula$^{4}$}\vspace{0.4cm}
\\
$^{1}$Departament de F\'\i sica Qu\`antica i Astrof\'\i sica and Institut de Ci\`encies del Cosmos (ICCUB), Universitat de Barcelona.\\ 
\ C.\ Mart\'{\i} i Franqu\`es, 1; E--08028~Barcelona, Spain\\
$^{2}$Instituto de Astrof\'\i sica de Andaluc\'\i a, IAA--CSIC. Glorieta de la Astronom\'\i a, s/n; E--18008~Granada, Spain\\
$^{3}$Departament d'Enginyeria Inform\`atica i Matem\`atiques, Universitat Rovira i Virgili. Av.\ Pa\"\i sos Catalans, 26;\\
\ E--43007~Tarragona, Spain\\
$^{4}$Aix Marseille Universit\'e, CNRS, Laboratoire d'Astrophysique de Marseille, UMR 7326. 13388 Marseille 13, France
}

\date{Accepted 2016 May 26. Received  2016 May 26; in original form 2015 September 01}

\pubyear{2016}

\begin{document}
\label{firstpage}
\pagerange{\pageref{firstpage}--\pageref{lastpage}}
\maketitle

\begin{abstract} 

We have developed a computationally competitive $N$-body model of a
previrialized aggregation of galaxies in a flat $\Lambda$CDM universe
to assess the role of the multiple mergers that take place during the
formation stage of such systems in the configuration of the remnants
assembled at their centres. An analysis of a suite of 48 simulations
of low-mass forming groups ($M_{\rm tot,gr}\sim 10^{13}\;h^{-1}\msun$)
demonstrates that the gravitational dynamics involved in their
hierarchical collapse is capable of creating realistic first-ranked
galaxies without the aid of dissipative processes. Our simulations
indicate that the brightest group galaxies (BGGs) constitute a
distinct population from other group members, sketching a scenario in
which the assembly path of these objects is dictated largely by the
formation of their host system. We detect significant differences in
the distribution of S\'ersic indices and total magnitudes, as well as
a luminosity gap between BGGs and the next brightest galaxy that is
positively correlated with the total luminosity of the parent
group. Such gaps arise from both the grow of BGGs at the expense of
lesser companions and the decrease in the relevance of second-ranked
objects in equal measure. This results in a dearth of
intermediate-mass galaxies which explains the characteristic central
dip detected in their luminosity functions in dynamically young galaxy
aggregations. The fact that the basic global properties of our BGGs
define a thin mass fundamental plane strikingly similar to that
followed giant early-type galaxies in the local universe reinforces
confidence in the results obtained.

\end{abstract}




\section{Introduction}\label{goals}

The brightest group galaxies (BGGs), as well as their cluster
counterparts (BCGs; throughout the paper, we use both terms
interchangeably), are among the largest known single structures in the
universe. They are mostly central passive giant elliptical (gE)
galaxies more luminous (and massive) than the average, with NIR
luminosities up to $\sim 12\,\lcIR$\footnote{$\lcIR \simeq
  \mbox{few}\times 10^{10} h^2 L_{{\rm NIR},\odot}$ is the
  characteristic luminosity marking the 'knee' of any NIR-band
  \citet{Sch76} luminosity function (LF) representative of the old
  stellar population.}, typical 1D central velocity dispersions
$\sigma\sim 150$--350\ \kms, and very little rotational support. As
with prototypical early-type galaxies (ETGs), their light profile is
well described by a de Vaucouleurs $R^{1/4}$ law over a large range
in radii. Some BGGs/BCGs are classified as D (gE possessing a large,
diffuse envelope) or cD (extra-large D) galaxies. The latter are
extremely large elliptical galaxies sitting near the centres of some
rich galaxy clusters which are surrounded by a distinct dynamical
stellar subsystem in the form of a very extended low-surface
brightness envelope of excess light over and above the $R^{1/4}$
profile defined by the inner regions made of old stars
\citep*{Kor89}. cDs are purely a rich cluster phenomenon, unlike fossil
groups \citep{Pon94}, which are isolated and likely relaxed
early-formed galaxy aggregations dominated by a central object,
sometimes as bright as a rich-cluster cD, surrounded by substantial
numbers of lower-luminosity satellites (at least 2 or more magnitudes
fainter by definition), all embedded in a common extended halo of
X-ray emitting gas the size of a galaxy group. They are thought to
result from the extensive merging of all the galaxies contained within
a small group, with the extended X-ray halo providing strong evidence
for the group origin. This has lead some to theorize that cD clusters
may result from the creation first of a fossil group and then the
accumulation via secondary infall of new galaxies around the fossil
remnant \citep[e.g.][]{Jon03}.

The observed differences in the total luminosity and luminosity
profiles with respect to standard ETGs \citep[e.g.][]{Sch86,BB01}
suggest that the formation histories of first-ranked galaxies follow a
pattern similar to that of their fainter peers only at high redshift
\citep[e.g.][]{DeLB07}. At early times, cooling flows are probably the
main fuel for the mass-growth of the bulk of the galaxy population. But
for galaxies residing in dense environments this source of growth is
removed as soon as they are incorporated into a bigger structure. In a
cosmological hierarchy, small groups of galaxies tend to form prior to
more massive structures. The gravitational collapse that precedes the
virialization of these precursor subunits leads to frequent, multiple
encounters among their member galaxies at relatively speeds comparable
to their internal stellar motions. When two or more galaxies collide
under such conditions, dynamical friction combined with strong
time-dependent mutual tidal forces redistribute the ordered orbital
kinetic energy into internal random energy, allowing the galaxies to
merge, in the absence of important energy dissipation, into a
velocity-dispersion-supported, ellipsoidal-looking system
\citep[e.g.][]{Bar88,Her92,BH92}. If this galaxy merging is not
accompanied by significant star formation, it should propitiate the
creation of massive and red early-type remnants in regions that later
may become part of a bigger structure.

It is critical for the substantial growth of BGGs that merging
develops in a local environment dynamically more mature than its
surroundings. The effectiveness of galaxy assembly in group-sized
units is largely reduced as soon as the latter start to coalesce into
a larger structure, deepening the overall gravitational potential and
increasing the relative velocity of galaxy encounters to values
($\gtrsim 500$\ \kms) well above the typical orbital speeds of their
stars \citep{Kun02}. In addition, the truncation of the radii of
non-central galaxies (satellites) surviving the virialization of a
protostructure may lead to time-scales for dynamical friction longer
than a Hubble time, so that any subsequent growth of the first-ranked
object through the cannibalization of its smaller companions will
likely be very modest, as the analytical accretion rate model by
\citet{Mer85} predicts. Indeed, once a galaxy aggregation relaxes, any
ensuing secondary infall will act to reduce the magnitude gap between
the highest ranked galaxy and the other system members that survive
the virialization stage \citep{LS06,vBB08}. This self-regulated growth
scenario in which the dominance of first-ranked galaxies depends on
the specific way cosmological infall develops around them allows to
reconcile the dependence of the observed properties of these objects
on the local environment with their standard candle nature
\citetext{\citealt{San72,PL95}; see also \citealt{DeLB07}}.

A BGG formation framework driven mainly by merging has to overcome an
important difficulty that is extensive to the general elliptical
population. The root of the problem lies in the combination of the
expectation that gas-rich late-type galaxies (LTGs) should account for
a large proportion of ETG progenitors with the fact that the relations
between the three major physical parameters informing about the size,
internal velocity, and scale of a galaxy, i.e.\ its mass -- usually
substituted by NIR luminosity, a directly measurable proxy, defining
what we hereafter refer as the $RVM$ or $RVL$ space --, differ
significantly\footnote{For a concise description of a galaxy's overall
  physical state, the scale should be complemented at least by a
  second independent property, such as colour or luminosity-weighted
  stellar age, providing condensed information on the star formation
  history. In this paper we work directly in $RVM$ space and thus
  avoid possible biases arising from luminosity weighting of stellar
  populations.}. As stated in \citet*{HCH08}, the observed
  differences in the properties of early- and late-type objects can be
  summed up in the different behaviour of the ratio of total
  (dynamical) mass to stellar mass, $M_{\rm tot}/\ms$, calculated
  within a fiducial radius (usually the radius at which half of the
  total light of the system is emitted). Thus, while the most massive
  discs and ellipticals have similar $M_{\rm tot}/\ms$ ratios, discs
  with masses below that of the Milky Way (MW) are substantially more
  dark matter dominated \citep[see, e.g.\ Fig.\ 1 in][]{CD15}. This
  means that the central phase space density of stars in intermediate
  and low-mass LTGs is systematically lower than in the cores of their
  elliptical counterparts. Since purely dissipationless merging cannot
  increase the density of the phase space \citep{Car86}, this
  difference is a major challenge for models that posit that
  ellipticals form through the merger of disc galaxies. The proposed
  solutions include merging progenitors that either are denser than
local discs \citep{AV05}, or harbour a compact central stellar region
\citetext{such as a bulge; \citealt{Her93}}, or alternatively, contain
a non-conservative component able to raise the inner stellar
concentration of the remnants as in \citet{Rob06}. These latter
authors used simulated wet (i.e.\ gas-rich) binary mergers of disc
galaxies with varying gas fractions to show that merger-induced
starbursts allow for an increase in central stellar density and
therefore a decrease in central $M_{\rm tot}/\ms$, thereby
demonstrating that gas dissipation can offer a viable explanation for
the observed tilt of the fundamental plane \citetext{FP;
  \citealt{DD87,Dre87}} of ETGs. However, as shown by \citet{Nov08},
while binary mergers of substantially gas-rich disc galaxies can
result in remnants with properties similar to those observed in the
low-mass, fast rotating ellipticals that comprise $\sim 80\%$ of the
galaxies detected in detailed ETG surveys such as SAURON
\citep{Ems07}, they have difficulties in producing objects comparable
to the massive, nearly spherical, non-rotating systems that make up
the remaining $20\%$ of the elliptical population \citep*[see
  also][]{MvBW10}. For this latter type of galaxies \citeauthor{Nov08}
finds that in order to get remnants with the right structural and
dynamical properties the key lies largely in dealing with realistic
assembly histories, which essentially means a non-regular sequence of
mergers with progressively decreasing mass ratios \citep[see
  also][]{Ose12,Moo14}. What is more, by using the zoom-in technique
for re-simulating haloes extracted from a full cosmological
simulation, he finds that in runs where the galaxies' gas supply is
continuously replenished the massive remnants that form tend to be
fast rotators. This suggests that the presence of a dissipational
component in a merger helps to maintain some coherent rotational
motion in the tightly bound region of the remnant by reducing the
efficiency of dynamical friction in transferring the progenitors'
orbital spin to the extended outer dark halo
\citep{Cox06}\footnote{Interestingly, \citet{Cev15} have recently
  shown that high-redshift massive galaxies subject to intense gas
  in-streaming are prone to suffer violent disc instabilities that
  contribute to the growth of compact central bulges with classical
  profiles by continuous transfer of gas-rich material from the disc
  to the spheroid.}. All these findings are specially relevant for the
most luminous ETGs which are known to possess smaller velocity
dispersions, larger sizes and fainter surface brightnesses than
expected if they have followed the same formation histories of their
less-massive counterparts \citep{HB09,MenA12}. It appears therefore
that a realistic hierarchy of multiple, essentially dry, mergers is
key in the assembly and evolution of these objects
\citep*{She03,NKB06,Des07,Ber07,Liu08,MenA12}. The contribution of
dissipation to the growth of first-ranked galaxies (by cooling flows
and/or wet mergers) must be then necessarily limited to early epochs
when galaxies were on average more gas-rich, the current large sizes
of BGGs being mostly the result of the non-dissipative capture of
numerous neighbouring objects facilitated by the physical conditions
of the environment in which they reside. For the same reasons, dry
hierarchical merging should also play an important role in explaining
the origin of the FP and related 2D scaling laws displayed by these
extreme ETGs.

A recent work by \citet*{TDY13,TDY15} constitutes the last, and
probably, also the most elaborate investigation to date of the
gravitational effects associated with extensive merging on the
internal structure and kinematics of central remnants in galaxy
associations. As stated by these authors, pure cosmological
simulations, which incorporate hierarchical merging in a
self-consistent manner, have been hitherto a resource scarcely used in
studies of the physical origin of the basic properties of first-ranked
objects in group/cluster-sized galaxy aggregations and the tight
scaling relations connecting them
\citep*[e.g.][]{SDTS04,NJO09,RS09,Ose12}. The most likely reason is
the high computational cost demanded by the large simulation sets that
this sort of statistical investigations require. Instead, merger
hierarchies, in both dry and wet simulations, are systematically
recreated through two or three non-overlapping steps of binary major
collisions of homologous systems that use the remnants of the previous
stage to realize the initial conditions of the next
\citep{Dan03,Rob06,Nov08,Nip09,Moo14}. These idealizations of multiple
merging, however, can hardly capture the diversity of galaxy
interactions, especially when it comes to the distribution of orbits,
frequencies and mass ratios, that take place in the course of cosmic
evolution. Another important drawback of this 'standard approach' to
hierarchical merging is that it does not allow as much orbital energy
transfer into the remnants as when the colliding galaxies are part of
a larger bound system. By contrast, \citeauthor{TDY13} follow the
assembly of BGGs from the outcome of over a hundred simulations of
groups of three to thirty gas-free spirals collapsing from
turnaround. Their progenitor galaxies, which have luminosities
randomly sampled from a wide range of a Schechter LF, are all
re-scaled versions of a large-bulge-fraction model of
M31. \citet{TDY13} place the most luminous object of each run at the
group centre and assign random locations to the remaining group
members up to a maximum radius equal to twice the $200\rhoc$ virial
radius the group would have at the initial redshift of the simulation
($z=2$). The spatial distribution of galaxies is such that their
extended NFW \citep*{NFW97} dark haloes form an essentially continuous
background of dark matter. The satellite galaxies are also given
random peculiar velocities biased both inward and radially to ensure
collapse. With this substantially more realistic pathway to account
for the importance of hierarchical merging in galaxy formation,
\citet{TDY15} are able to show that multiple dry mergers of spiral
galaxies produce central dominant remnants lying along a tight FP that
is tilted, albeit somewhat less than suggested by
observations. Therefore, their findings lend credence to the idea that
in high-density environments gravitational processes on their own can
be largely responsible for the formation of first-ranked ETGs, thus
obviating the need for the intervention of progenitors with abundant
($\gtrsim 40\%$) gas fractions, as suggested for instance by
\citet{Rob06} and \citet{HCH08}.

The present work, which follows the line initiated by the numerical
experiments of e.g.\ \citet{Dub98}, \citet*{AGGG01}, \citet{Dan03},
\citet{Gao04}, \citet{DeLB07}, \citet{vBB08} or \citet{RS09},
represents further progress in the investigation of the role of pure
collisionless dynamics in the assembly of the largest spheroids in the
universe. Specifically, we carry out a study of the optical properties
of our simulated groups and BGGs that complements the work done by
some of us assessing the impact of dissipationless hierarchical
merging on the mass FP \citetext{\citealt{PS16}; hereafter
  \citeauthor*{PS16}}. The findings of both investigations provide
conclusive evidence in favour of the fundamental contribution of dry
multiple merging in small collapsing galaxy aggregations to the
formation of realistic first-ranked galaxies. As in
\citeauthor{TDY13}'s papers, we will use for this purpose controlled
high-resolution $N$-body simulations of the virialization phase of
small galaxy groups, but with a set-up yet somewhat further elaborated
in certain aspects. The main differences in our approach are the
simultaneous use of late- and early-type galaxy models intended to
account for the expected mix of progenitors' populations, the
non-homologous scaling of the progenitors' properties, the inclusion
of a common dark matter background not associated with any particular
galaxy, and the scaling of the mass of the progenitors' dark haloes to
values appropriate for the initial redshift of the simulation,
$\zi=3$, which is also significantly larger. The focus is to build
controlled simulations of group formation that approximate as closely
as possible to pure cosmological experiments. By contrast, we are more
restrictive than \citeauthor{TDY13} in terms of the range of galaxy
numbers and total masses of the simulated groups, which we have fixed
to specific typical values (see next Section), without this
undermining the conclusions of the present investigation. Analysis of
simulations of galaxy aggregations of higher mass and broader
membership is currently underway (Perea \& Solanes 2016, in
preparation). The new simulations will allow us to investigate
possible correlations between these two factors and the main
properties of BGGs.

The paper begins by explaining in detail the approach followed to
create the simulated BGGs (Sec.\ \ref{simulations}), as well as the
methodology used to analyse them (Sec.\ \ref{analysis}).
Section~\ref{optical} focuses on the imprint left by group formation
on the optical properties of their central remnants. Next, we examine
in Section~\ref{scalings} the joint distribution of global properties
in the 3-space of virial-related variables and check its consistency
with real data. Finally, in Section~\ref{conclusions} we discuss the
results of this work and comment on its implications on theories of
ETG formation. We also include three appendixes:
Appendix~\ref{num_resol} evaluates the effects of numerical heating in
our galaxy models; Appendix~\ref{halo_evolution}, describes the model
adopted to account for the secular growth of galaxy dark haloes; and
Appendix~\ref{deneval}, which provides details about the procedure
used to elaborate our surface density maps. Both the forming group and
the galaxy models are build according to the predictions of the
standard concordance flat $\Lambda$ cold dark matter ($\Lambda$CDM)
cosmology that we use throughout the paper. Specifically, we adopt a
matter density parameter $\Omo=1-\Olo=0.26$, and a reduced Hubble
constant $h=H_0/(100\ {\mathrm{km/s/Mpc}})=0.72$ from the WMAP 5-year
results \citep*{Kom09}.

\section{Galaxy group simulations}\label{simulations}

Our group simulations are designed to study galaxy aggregations that
are in the act of formation. It is in this type of environment in
which it is expected that the contribution of gravitational dynamics
have the greatest impact on the evolution of galaxies and
intergalactic medium. More specifically, we aim to reproduce the
properties of relatively compact low-mass ($M_{\rm tot,gr}\sim
10^{13}\;h^{-1}\msun$) galaxy associations, with typically a few
bright ($L\geq 0.5\;\lc$) members, small line-of-sight (l.o.s.)
velocity dispersions ($\sigma_{\rm{los}}\sim\mbox{few} \times
10^2$~\kms), and low X-ray luminosities ($L_X <
10^{42}\;\mbox{erg}\;\mbox{s}^{-1}$), that have not yet
virialized. According to the results of the $\Lambda$CDM Millennium
Simulation \citep{Spr+05}, the mean formation
redshift\footnote{Defined as the first redshift slice in which the
  largest progenitor of a dark halo has half the mass of the final
  halo.} of overdensities this size is $1.0\pm 0.5$
\citetext{\citealt*{McBr09,Har06}; see also \citealt*{Gio07}}. Thus,
in the local universe, the groups at the centre of our focus would be
among the dynamically youngest systems of their class, reaching
turnaround at cosmic times ($z\sim 1$) not too different from the
typical virialization epoch of these structures.

\subsection{Group model}\label{grp_model}

Our groups are created as isolated overdensities that first expand
linearly, then turn\-around, and finally undergo a completely
non-linear collapse, as done in the pioneering works by \citet*{DGR94}
and \citet*{GTC96}. Each simulated group starts being a (nearly)
uniform spherical primordial fluctuation at redshift $\zi=3$ (about
one-eighth of the present time $t_0$). At this cosmic time,
galaxy-sized haloes have already acquired most of its mass
\citep{Li07,Ste08}, but their cold baryonic cores need not necessarily
be as developed. The star formation rate of a non-negligible fraction
\citetext{at least $30\%$ according to \citealt*{Sar06}} of the
brightest galaxies may begin to shut down abruptly at $z\sim 3$
\citep[e.g.][]{Cat06}, indicating that these objects are nearly fully
grown at the chosen $\zi$. The smaller representatives of the galaxy
population, however, may endure most of their stellar mass growth
at lower redshifts -- the well known 'downsizing' effect --, a fact
that we have decided to ignore (we justify this choice in Section
\ref{stellar_core}).

\begin{figure}
\centering
\includegraphics[width=\linewidth,angle=0]{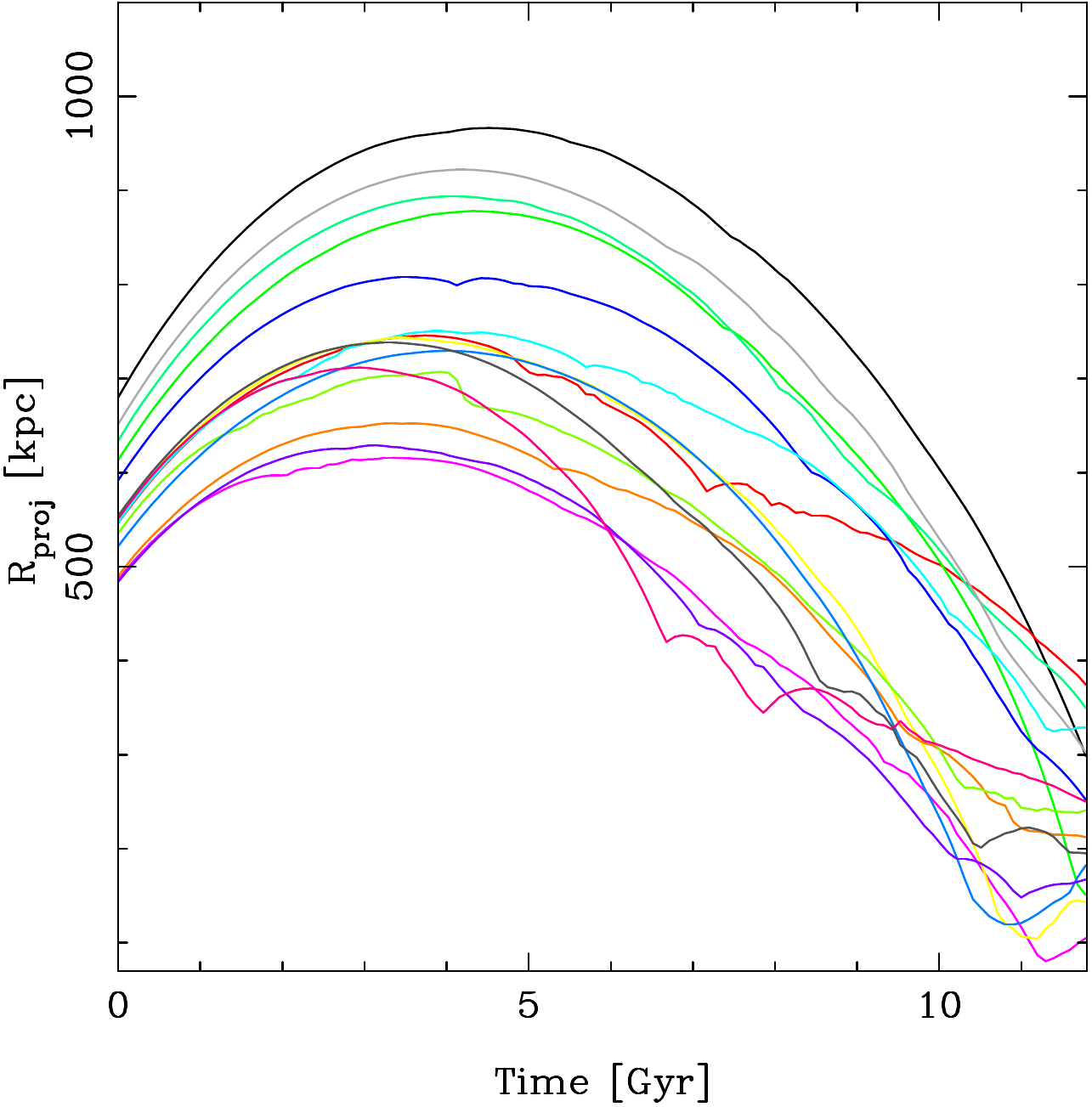} 
\caption{\small Secular evolution of the mean projected inter-galactic separations $R_{\rm proj}=\langle d_{ij}\rangle$ in the $XY$ plane for a representative subset of fifteen of our 48 simulated groups. All our simulations start at $\zi = 3$, reach turnaround at $z_{\rm turn}\sim 1$, and are evolved until the present epoch when most groups are near the transition between the collapse and rebound phases and $R_{\rm proj}$ is about its absolute minimum.}\label{fig_1}
\end{figure}

The value of the initial overdensity of our mock groups, $\deli$, in
which only the growing mode is present, i.e. $\delta(t)=(3/5)\deli
(t/\ti)^{2/3}$, is chosen so that a perfectly homogeneous top-hat
perturbation would collapse at $z=0$, which is also the approximate
redshift $\zf$ at which we end our simulations. The initial positions
of the background particles (including the centre of mass of the
galactic haloes of member galaxies) are randomly distributed inside the
group volume, while their initial velocities are purely radial,
following the inner perturbed Hubble flow \citep{Ber85}
\begin{equation}
v(r)=(1-\deli /3)H_{\rm i}r\;,
\end{equation}
where $v(r)$ is the radial expansion velocity at a distance $r$ from
the group centre, and $H_{\rm i}$ is the unperturbed Hubble parameter
of the universe at $\ti\equiv t(\zi)$. In our experiments, the time
elapsed before turnaround, about 1/3 of the total simulation time (see
Fig.~\ref{fig_1}), which marks the onset of the non-linear phase,
serves for gravitational interactions between group members to build
up peculiar velocities and $n$-point correlations among their
positions, thus guaranteeing a reasonable approximation to a
conventional galaxy group before merging becomes
widespread. Nevertheless, it must be noted that this early (mild)
clustering is only relevant for the (very) few mergers that take place
before or around the turnaround epoch. Since the initial state of our
groups is far from equilibrium, in the remaining 2/3 of the total
simulation time, which follows the gravitational collapse of the
system, galaxies move in a mean gravitational field that becomes
widely fluctuating and increasingly erratic as the violent relaxation
of the system they inhabit starts to unfold. As a result, the outcome
of the mergers that take place in this highly non-linear phase gets
largely detached from the initial orbital anisotropy of progenitor
galaxies.


\begin{figure*}
\begin{minipage}{175mm}
\centering
\includegraphics[width=\linewidth,angle=0]{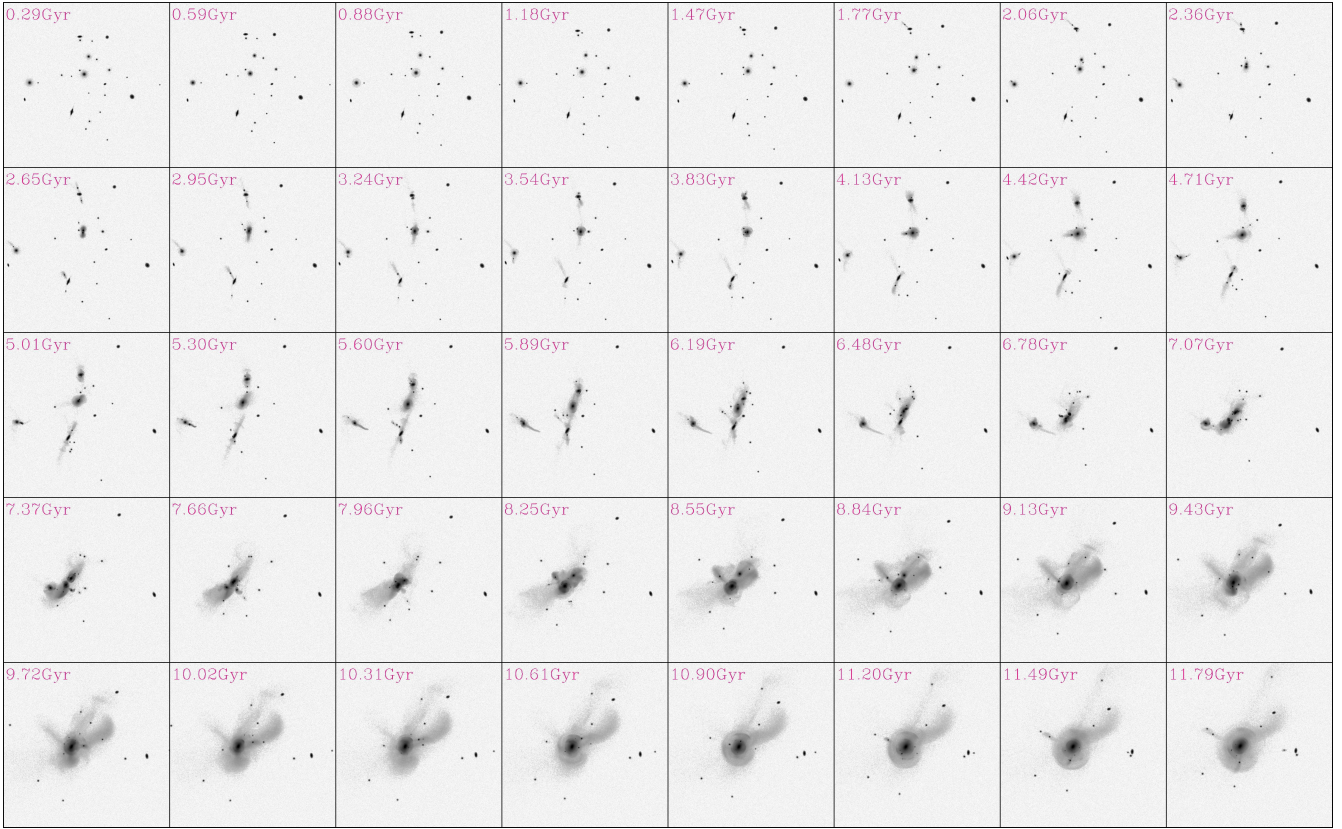}
\caption{\small Series of 40 logarithmic surface density maps uniformly spaced in time showing the evolution during the previrialization stage of the luminous matter of group \#8336. The darker the grey tones in the snapshots the higher the density of particles. Our experiments use 5 million particles initially distributed among 25 galactic haloes with virial masses drawn from a \citeauthor{Sch76} MF with $M/\mc\geq 0.05$ and $\alpha = -1$, and a uniform common background of dark matter. The accumulated simulation time is listed in the top-left corner of each image. In the last Gyr of evolution this particular group resembles a typical fossil galaxy.}\label{fig_2}
\end{minipage}
\end{figure*}

\begin{figure*}
\begin{minipage}{175mm}
\centering
\includegraphics[width=\linewidth,angle=0]{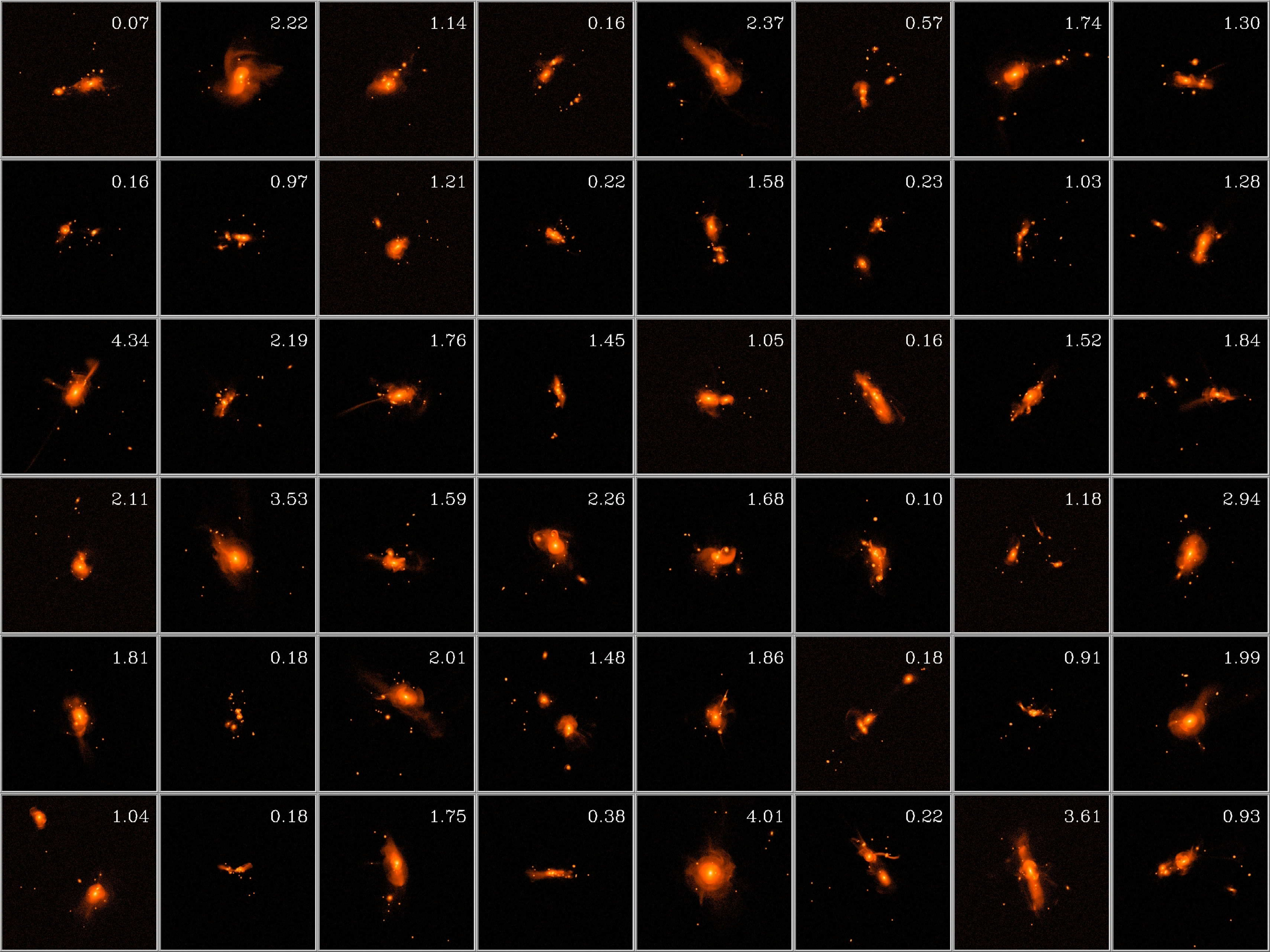}
\caption{\small Final snapshots of our 48 simulated groups. The
  lighter the colour the higher the density of
  particles. This plot illustrates the enormous variety of final
  configurations produced by our experiments, as there are virtually
  no two groups alike. In some cases, first ranked galaxies remain
  largely unchanged over the entire simulation, while in others we
  observe the creation of a large, dominant BGG as the result of
  multiple mergers. The figure in the upper-right corner of each panel
  informs about the size of the magnitude gap $\Delta\magn_{12}$ (see
  \S~\ref{m12}). Also note that practically all images show the
  presence of intragroup debris in various forms (extended low surface
  brightness features, shells, narrow streams, plume- and
  umbrella-shaped structures,$\ldots$) which has been ripped from the
  galaxies mostly by the strong gravitational interactions that take
  place during the highly nonlinear collapse of their parent
  group. Unless stated otherwise the values of the variables shown in
  this and following figures correspond to the redshift of observation 
  $\zf\simeq 0$.}\label{fig_3}
\end{minipage}
\end{figure*}

The initial radius of each simulated group is estimated from the
dissipationless spherical collapse model. Assuming that pressure
gradients are negligible, a top-hat perturbation of amplitude
$\deli>0$ at $\ti$ evolves like a Friedmann universe whose initial
density parameter is given by \citep[cf.][Eq.~(14.1.2)]{CL02}
\begin{equation}\label{denparam}
\Op(\ti)=\frac{\rho_{\rm p}(\ti)}{\rhoc(\ti)}=\frac{\rho(\ti)(1+\deli)}{\rhoc(\ti)}=\Om(\ti)(1+\deli)\;,
\end{equation}
where the suffix 'p' identifies quantities associated with the
perturbation, while $\rho(\ti)$, $\rhoc(\ti)$, and $\Om(\ti)$ refer
to the unperturbed background cosmology. Since dark energy starts to
dominate the dynamics of the universe near the present time
($z\lesssim 0.6$) and the condition for the collapse of the
perturbation is $\Op (\ti)>1$, it is safe to neglect the contribution
of $\Lambda$ to its evolution. Under this approximation the expansion
of the perturbation is described by the equation
\begin{equation}\label{pevol}
\left(\frac{\dot{a}}{a}\right)=H_i^2\left[\Op (t_i)\frac{a_i}{a}+1-\Op (t_i)\right]\;,
\end{equation}
with $a$ the cosmic scale factor.

When the perturbation turns around at a time $t_{\rm{ta}}$, $\dot{a}=0$, so
\begin{equation}
\rho_{\rm p}(t_{\rm{ta}})=\frac{3\pi}{32Gt_{\rm{ta}}^2}\;,
\end{equation}
where, because of the symmetric evolution of the perturbation (even when 
$\Ol\neq 0$), $t_{\rm m}$ is equal to one half the adopted collapse
time $\tf=t(\zf)$.

Once $\rho_{\rm p}(t_{\rm{ta}})$ has been determined, the initial
overdensity of the perturbation representing the group can be
calculated from the implicit equation
\begin{equation}
\frac{\rho_{\rm p}(t_{\rm{ta}})}{\rhoc(\ti)}=\Om(\ti)(1+\deli)\left[1-\frac{(1-\deli/3)^2}{\Om(\ti)(1+\deli)}\right]^3\;,
\end{equation}
and then, with the aid of equation~(\ref{denparam}), one finally obtains
the initial radius of the group
\begin{equation}
R_{\rm p}(\ti)=\left[\frac{3\Mp}{4\pi\rhoc(\ti)\Om(\ti)(1+\deli)}\right]^{1/3}\;,
\end{equation}
where for the present work we have fixed the total mass (dark and
luminous) of the perturbations associated with all our simulated
groups to the value $\Mp=M_{\rm{tot,gr}}=10^{13}\,h^{-1}\msun$.

\subsection{Galaxy models}\label{gal_models}

Each group region initially contains $\Ngal$ non-overlapping extended
galaxy haloes whose structural and dynamical properties (total mass,
spin, concentration, and density profile of the dark component) set
the scalings of their central baryonic (stellar) cores. The different
galactic components are in dynamical equilibrium, representing a
stationary solution of the relevant dynamical equations for
collisionless matter. As shown below, our galaxy models, which by-pass
all the issues of ab-initio galaxy formation, produce objects with
properties that are both consistent with the hierarchical build-up of
these systems and well-motivated observationally.

\subsubsection{Dark matter haloes}\label{dark_halo}

The total masses of individual galaxy haloes, $\Mvir$, are randomly
drawn from a Schechter-like function of asymptotic slope $\alpha=-1.0$
and characteristic mass $\mc=\Mtwel\equiv 10^{12}\;h^{-1}\msun$. We
set a lower limit in mass of $0.05\;\mc$, as we do not expect smaller
haloes to play a significant role in the results. We note that, at
this point, the values $\Mvir$ represent the virial masses of the
progenitor galaxies at the redshift ($\zf\sim 0$) of observation. The
total galaxy masses and taper radius of the dark haloes will be scaled
down later in Sec.~\ref{simul_set-up} to values consistent with the
initial redshift ($\zi=3$) of the simulations.

For a given $\Mvir$, the remaining global parameters of the haloes are
fully determined by the background cosmology (i.e., are independent of
the assumed halo structure) given a time of observation. Cosmological
$N$-body simulations show that the virial radius $\Rvir$ of a CDM halo
of mass $\Mvir$ observed at a redshift $z$ can be defined as the
radius at which the mean halo mass density is the mean density of the
universe $\rhou$ times the virial overdensity $\Dvir$ of a collapsed
object in the top-hat collapse model at that redshift. Thus, we have:
\begin{equation}\label{Rvir}
\begin{split}
\Rvir=1.63\times 10^{2} & \left[\frac{\Mvir}{\Mtwel}\right]^{1/3}\cdot\\ & \left[\frac{\Omo\Dvir(z)}{200}\right]^{-1/3}(1+z)^{-1}\,h^{-1}\,{\mathrm{kpc}}\;,
\end{split}
\end{equation}
where for the family of flat ($\Om+\Ol=0$) cosmologies \citep[cf.][]{BN98}
\begin{equation}\label{Dvir}
\Dvir(z)\simeq\{18\pi^2+82[\Om(z)-1]-39[\Om(z)-1]^2\}/\Om(z)\;,
\end{equation}
with
\begin{equation}\label{Omz}
\Om(z)=\frac{1}{1+(\Olo/\Omo)(1+z)^{-3}}\;,
\end{equation}
which results in $\Dvir\simeq 368$ at $z=0$ for the adopted
cosmological parameters.

Similarly, the halo circular velocity $\Vvir$ (i.e.\ the circular velocity at
$\Rvir$) is:
\begin{equation}\label{Vvir}
\Vvir=\left[\frac{\Rvir}{h^{-1}\,{\mathrm{kpc}}}\right]\left[\frac{\Omo\Dvir(z)}{200}\right]^{1/2}(1+z)^{3/2}\,\mbox{km\ s}^{-1}\;, 
\end{equation}
while the dynamical time for all haloes, a quantity that depends only
on the world model and redshift chosen, is defined as
\begin{equation}\label{Tdyn}
\begin{split}
T_{\rm{dyn}} \equiv & \frac{\Rvir^{3/2}}{(G\Mvir)^{1/2}} = \\ & \ \ 0.978 \left[\frac{\Omo\Dvir(z)}{200}\right]^{-1/2}(1+z)^{-3/2}\,h^{-1}\;{\mathrm{Gyr}}\;.
\end{split}
\end{equation}
The above relationships are useful to compare our simulations with
real galaxies as long as we fix the equivalence between simulation and
physical units. We do this by taking the unit mass in our simulations
equal to the characteristic mass adopted for the mass function (MF) of
galaxies, so $\msu=\mc=\Mtwel$. This is also the total halo mass
commonly associated with a MW-sized galaxy in the local universe
\citetext{see, e.g., \citealt*{KZS02,B-K09}, and references
  therein}. The corresponding scalings for length, velocity and time
for $z=0$ are $\rsu = 208\,h^{-1}\;{\rm{kpc}}$,
$\vsu=144~\mbox{km\ s}^{-1}$ and $\tsu=1.41\,h^{-1}\;{\rm{Gyr}}$,
respectively.

We model the inner structure of the galactic dark matter haloes using
a NFW density profile of virial radius $\Rvir$. This latter parameter
is inferred from equation~(\ref{Rvir}), taking into account that the
selected values of $\Mvir$ correspond to the redshift of
observation $\zf\simeq 0$. The halo concentration, $c\equiv\Rvir/\rs$,
used to determine the scale radius $\rs$ of the NFW profile, is
inferred from the median $M$--$c$ relation for the standard concordant
$\Lambda$CDM cosmology found in high-resolution $N$-body simulations,
which over the range of masses we are interested, $10^{10}\lesssim
\Mvir/(h^{-1}\msun) \lesssim 5\times 10^{12}$, is well approximated by
the equation
\begin{equation}\label{M-c}
c=9.35\left[\frac{\Mvir}{\Mtwel}\right]^{-0.094}(1+z)^{-1}\;,
\end{equation}
where the adopted redshift dependence corresponds to the scaling
$c(a)\propto a=(1+z)^{-1}$ of the median concentration of haloes of
fixed mass found in \citet{Bul01} and \citet{Wec02} simulations
\citetext{see also \citealt{MDvB08}}. 

A neat rotation consistent with the adoption of a non-null value for
the dimensionless internal spin parameter $\lambda$ (see
\S~\ref{stellar_core}) is imparted to the dark matter haloes to
account for the contribution of the gravitational tidal torques that
may have arisen in the course of the growth of the initial
perturbations. Inspired by \citet{SW99}, the net fraction of halo
particles required to produce a given $\lambda$ is taken to be
proportional to the ratio between the streaming velocity of the
galactic halo and the local azimuthal circular velocity. Since it is
reasonable to expect that the dark haloes of ETGs have generally a
lower spin than those hosting late-type objects \citep{Hua12}, we
reduce by half the value of $\lambda$ for the haloes hosting pure
spheroidal stellar distributions. The orientations of the galactic
spins within the group are selected at random.

\subsubsection{Baryon cores}\label{stellar_core}

A $5\%$ of the total mass of each galaxy is placed in a stellar core
following either a disc-bulge or a pure spheroidal distribution. For
galaxies with masses equal to or larger than $0.1\msu$ morphologies
are established using a Monte Carlo technique that assumes a
mass-independent LTG fraction of 0.7, typical of the field \citep[see,
  e.g.\ Fig.\ 9 in][]{Pos05}. Galaxies below this mass threshold,
which in our simulations contribute around $5\%$ of the total stellar
mass, are assumed to host only spheroidal stellar distributions. This
setup, which does not reproduce results on the morphological
mix of local galaxies with masses similar to those of the two
Magellanic Clouds \citep{Kel14}, is actually a matter of simplicity
since such small galaxies are marginally resolved in our simulations
and therefore highly sensitive to numerical heating (see
App.~\ref{num_resol}). Besides, as we will show in
Section~\ref{starMF}, objects of this size survive practically
unscathed the collapse of their host groups thanks to their reduced
internal velocities. Changing the inner structure of the smallest
galaxies included in our simulations would not alter the main results
of the present study.

At this point, it is important to stress that our LTG and ETG models
do not pretend to accurately reproduce the structural and dynamical
characteristics of the different kinds of galaxies that can be present
in a group. Instead, we aim to provide a sensible representation of
the two basic shapes of the stellar distributions expected to populate
the universe at all epochs. Namely, relatively extended and fragile
discy objects with a substantial rotational component, and
dynamically-hot spheroids, i.e.\ stellar systems that are stabilized
essentially by internal random motions and have centrally concentrated
ellipsoidal profiles.

To realize the late-type stellar distributions we adopt the usual 3D
exponential-sech-squared profile (Spitzer's isothermal sheet), which
in cylindrical coordinates is given by
\begin{equation}
\rho_{\rm{disc}}(R,z)=\frac{0.05\Mvir}{4\pi\Rd^2z_{\rm{d}}}\mbox{exp}(-R/\Rd)\mbox{sech}^2(z/z_{\rm{d}})\;,
\end{equation}
where we assume a disc scale-height $z_{\rm{d}}=0.2\Rd$, as it
is typical in high surface brightness galaxies in the local universe
\citep[e.g.][]{BK04}.

Disc scale-lengths, $\Rd$, are initialized according to the adaptation
by \citet{DS10} of the analytic formalism for disc formation from
\citet*{MMW98}, which relies on specific angular momentum conservation
during gas cooling and adiabatic halo response to gas inflow, and that
reproduces the (slopes and zero-offsets of the) main observed 2D
scaling relations for nearby discs. According to this model
$\Rd=k\cdot\Rvir$, with factor $k$ being a function, for a given halo
profile, of $c$, set by the total halo mass (Eq.~\ref{M-c}), and the
adopted $\lambda$ and stellar mass fraction $f_\star\equiv \ms/\Mvir$
values. For our model LTGs we take $\lambda=0.04$, the median value of
this parameter found in cosmological simulations \citep{Sha06}, which
when combined with $f_\star=0.05$ for galaxies embedded in NFW haloes,
produces discs obeying the observed local relation between the average
specific angular momentum, $\iota_{\rm {d}}=2\Rd V_{\rm max}$, and
the peak rotation velocity, $V_{\rm max}$, inferred from the width
of the global \hi\ profile. More details can be found in \citet{DS10}.

\begin{table*}
\begin{minipage}{105mm}
\scriptsize
\centering
\caption{Parameters common to all simulations}
\label{common_params}

\begin{tabular}{ll}
\hline \hline
Description & Value \\
\hline 
Total group mass $M_{\rm{tot,gr}}$ & $10\,\msu\sim 1.4\times 10^{13}\,\msun$\\
Initial redshift $\zi$ & $3.0$\\
Observation epoch $\zf\sm\zcoll$ & $\sim 0$\\
Look-back time since the beginning of simulation & $11.42$ Gyr\\
Initial overdensity of the group $\delta_i$ & $0.80$\\
Initial radius of the group & $0.91$ Mpc\\
Initial number of group member galaxies $\Ngal$ & 25\\
Asymptotic slope $\alpha$ of the Schechter galaxy mass function & $-1.0$\\
Low-mass cutoff of the Schechter galaxy mass function & $0.05\,\mc$\\
Fraction of large ($M\ge 0.1\msu$) LTGs & 0.7\\
Bulge-to-disc mass fraction for LTGs & 0.25\\
Total number of particles in a $\msu$ galactic halo & $500.000$\\
Initial fraction of stellar particles in a $\msu$ galactic halo & 0.6\\
Stellar particle mass & $\sim 2.3\times 10^5\,\msun$\\
DM particle mass & $\sim 6.6\times 10^6\,\msun$\\
DM-to-stellar particle mass fraction & 28.5\\
Background-to-galactic halo DM particle mass fraction & 1.0\\
Plummer-equivalent softening length for stellar particles & $30$ pc\\ 
Plummer-equivalent softening length for DM particles & $\sim 160$ pc\\ \hline
\end{tabular}
\end{minipage}
\end{table*}

Both bulges and ETGs are represented by zero-angular momentum stellar
spheroids supported by velocity anisotropy and obeying a spherical
\citet{Her90} mass density profile
\begin{equation}\label{ETG}
\rho(r)=\frac{\ms}{2\pi}\frac{\rH}{r(\rH+r)^3}\;,
\end{equation}
where $\rH$ is the profile scale length. Given that the applicability
to merger-made galaxies of the adiabatic-gas-inflow model for disc
formation is very questionable \citetext{but see
  \citealt{Pad04,Dut11}}, we have decided to create initial conditions
for ETGs using instead simple empirical laws adjusted to match the
mean scalings observed for this population in the local
universe. Thus, we relate the size measure of our elliptical galaxy
models represented by the effective radius of their projected
luminosity profiles, $\Reff$, to their total mass via the empirical
formula
\begin{equation}\label{empirical_re}
\Reff\approx 1.8153\;\rH =2.05\left[\frac{\Mvir}{\Mtwel}\right]^{5/8}\,h^{-1}\,{\rm{kpc}}\;,
\end{equation}
which matches the observed size-luminosity scaling for Sloan Digital
Sky Survey (SDSS) ETGs reported by \citet{Ber03b} in the $i^*$-band
(Table 1 of this paper) if one assumes $f_\star=0.05$, a fixed
rest-frame mass-to-light ratio $\Upsilon_I$ of $2.90\,h$ solar
units\footnote{This value is in reasonable agreement with the
  expectation for a typical nearby $\lc$ galaxy inferred from the
  $\Upsilon_I$ vs $L_I$ correlation derived by the SAURON project
  \citep{Cap06}.}, and a colour equation $(i^*-I)=0.53$ mag
\citep*{FSI95}. We note that Equation~(\ref{empirical_re}) is also
consistent with the $\Reff$--$\ms$ relation for massive galaxies
inferred by \citet{She03} likewise from SDSS data. For LTGs, we simply
add a bulge with a constant bulge-to-disc mass fraction of $1:4$,
typical of Sb galaxies \citep{Gra01}, and a bulge-to-disc size ratio
$\Reff/\Rd=0.20$.

Both spheroidal and exponential distributions are extended into the
radial direction up to a radius encompassing $95\%$ of the total
stellar mass. We have verified that in this manner we generate
galaxies with comparable minimum local stellar density at the outer
radius. We also note that a circular aperture of two Petrosian radii,
used to define the Petrosian magnitude of a galaxy, encloses about the
same fraction of the stellar light for the corresponding typical
S\'ersic profiles of indexes $n = 1$ (exponential) and $n = 4$ (de
Vaucouleurs), in the presence of seeing \citep{Bla01}. For LTGs, the
stellar disc distribution is also truncated in the vertical direction
at $10\;z_{\rm{d}}$.\footnote{This and the truncation in $R$ refer
  exclusively to the fact that we are tapering the stellar
  distributions at finite distances; in no case they imply the
  truncation of the adopted initial stellar masses.}

\subsection{\emph{N}-body realizations of galaxy groups}\label{simul_set-up}

We have run a total of 48 $\mathcal{O}(5\times 10^6)$-particle
simulations of galaxy groups containing initially $\Ngal=25$
galaxies. With this initialization, the average number of independent
galactic haloes that remain at the end of our runs is
consistent with the statistics of sub-halo abundance in typical
group-sized $\Lambda$CDM haloes \citep{Gao11,TDY13}, while
facilitating the generation of a sufficient number of large ($M\geq
0.5\;\mc$) discy progenitors from the adopted galaxy MF.

Our initial galaxy models have been built with the aid of a
much-expanded version of the computationally efficient
\magal\ open-source code \citep*{BKPG01} included in the
\nemo\ Stellar Dynamics Toolbox
(\texttt{http://bima.astro.umd.edu/nemo/}).  The backbone of
\magal\ is the program \bdgal\ \citep{Her93}, which allows one to
create either two-component (spheroid-halo) or three-component
(disc-bulge-halo) galaxies in near equilibrium using different halo
density profiles. Among other improvements, the upgraded code is now
fully portable and includes the possibility of adding extra components
to the galaxies, such as a second co-planar disc or a central
supermassive black hole, as well as scaling the output to arbitrary
units. More importantly, the refurbished program includes a more
accurate treatment of rotation in the centre of discs that eliminates
the annular disturbances generated by the original algorithm, enabling
the production of galaxies that can remain fully stable when evolving
in isolation for more than a Hubble time (see App.~\ref{num_resol}).

The total number of bodies per galaxy scales with its total mass. We
assign $500,000$ particles per $\msu$ in galactic haloes at $z=0$, 60
per cent of which are used to represent the central luminous core
(with this arrangement at least $\sim 40\%$ of the total number of
particles in our simulated groups are luminous). Taking into account
that we are modelling galaxies with a current baryon content of $5\%$,
the DM to stellar particle mass ratio is then 28.5. The stellar
Plummer equivalent softening length has been set to 30 pc. For the
dark component the softening length is scaled with the square root of
the body mass ratio, resulting in a value of $\sim 160$ pc. Thus, both
body types (light and dark) experience the same maximum force. We have
also taken the precaution of running in Appendix~\ref{num_resol}
long-term stability tests to quantify the breadth of the changes
suffered by the internal structure of the galaxies simulated with
these initial conditions, finding that the effects of numerical
heating are maintained within acceptable limits for the duration of
our experiments. Total energy conservation, which serves to evaluate
the accuracy of time integration, is better than $0.5\%$ in all cases.

As stated earlier, we have used the virial masses of the progenitor
galaxies selected at the observation epoch $\zf\simeq 0$ to set the
structural parameters of their central stellar cores. Now, we need to
scale down the masses and radii of the dark haloes to values
consistent with the initial redshift of the simulation $\zi=3$. This
is done by applying Equation~\ref{massincr} of the halo growth model
described in Appendix~\ref{halo_evolution}. The luminous components,
however, are left unchanged because in our simulations most mergers
take place at $z\lesssim 0.4$, an epoch in which galaxies are
relatively similar to those in the local volume. As shown by a large
number of studies \citep[][to name a
  few]{LaB03,TP05,Tru06,Kas12,H-C13}, the relative abundances, as well
as the luminosity- and stellar mass-size relations of galaxies with
both early- and late-type characteristics have not evolved
significantly since these moderate redshifts, at least for galaxies as
large as the MW, which are the main drivers of the interactions in our
simulations. Once all the galaxies of a group have been generated, we
fill evenly the empty space of the common volume with DM particles
(identical to those that make the galactic haloes) until a total group
mass (dark and luminous) of $10\;\msu$ is reached. This homogeneous
background is intended to represent the matter that exists in scales
below the smallest of our mock galaxies, which is expected to be
composed by a myriad of smaller-scale systems (with $M\lesssim 6\times
10^{10}\msun$), many of whom have likely collapsed prior to the epoch
at which our simulations begin, and for which we do not resolve
neither their inner structure nor their spatial distribution within
the group volume. The common dark haloes account at $\zi=3$ for 32 to
88 per cent of the total group mass, with a median value of $\sim 62$
per cent. This measure, its complementary initial fraction of mass in
bound subhaloes, and, especially, the associated and invariant total
amount of stellar mass in the system, are the quantities best suited
to characterize our groups, given that all of them have identical
total mass and initial richness.

Our groups are evolved for about 11.5 Gyr from $\zi=3$ to $\zf\simeq
0$ -- from $\ti=0$ to $\tf=6$ in the adopted simulation units -- using
the collisionless part of the public parallel code GADGET2
\citep{Spr05}, which employs a hybrid TreePM method to compute
gravitational forces. The runs require about $3,800$ hours of CPU each to
complete. We use adaptive time stepping, which for our simulations
leads to maximum time resolutions of the order of $\mbox{few}\times
10^{-5}$ simulation units, and save snapshots each $\Delta t =
0.05$. In Fig.~\ref{fig_2} we provide a composite of selected
snapshots equally spaced in time showing the evolution of one of our
galaxy groups.

Table~\ref{common_params} lists the most important parameters that all
of our runs have in common, while Fig.~\ref{fig_3} shows the final
snapshot of all our 48 simulated groups (unless stated otherwise
depicted group images correspond to orthographic projections
on to the $Z=0$ plane). This plot serves to illustrate the large
variety of collapsing group configurations that are produced in our
experiments. Although it is a subject that we expect to analyse in a
forthcoming publication, note the richly varied intergalactic light
structures that form too. In Appendix~\ref{deneval}, we provide the
details of the kernel density estimation technique used for generating
the density maps presented in this paper.

\section{Analysis methodology}\label{analysis}

The extraction of BGGs properties has followed a two-stage approach.

In the first place, we identify all the galaxies present in each of
the evenly spaced 120 snapshots stored from each group simulation and
derive both their total luminous masses and mean positions. Galaxy
identification is based on the projected distribution of the visible
particles, thus mimicking classical observational methods dealing with
the density of projected light. We rely on the SExtractor software
\citep{BA96} to perform this task. In a second stage, the information
on the number and positions of the galaxies is used to assign galaxy
membership to the luminous particles in the final snapshots and, after
ranking those belonging to the largest central galaxy in each of the
groups according to their galactocentric distance, to infer the
effective radius $\Reff$, surface mass density, $\SBe$, and projected
internal velocity dispersion, $\sige$, of our collection of mock BGGs.

Simulated groups are imaged along three independent
l.o.s.\ corresponding to the three Cartesian axes. At present, it has
not been deemed necessary to create additional projections since (1)
any other randomly oriented map can be obtained by a linear
combination of the principal images, and (2) in this paper we deal
mostly with global galaxy properties which, in general, change little
with orientation. Mock sky-images of the groups are created after
putting each group at a fiducial redshift $\zo=0.065$ representative
of the typical mean depth of BCG spectroscopic samples
\citep[e.g.][]{Liu08}. We then apply a pixel scale of $0.4$
arcsec/pixel, comparable to the SDSS scale ($0.396$ arcsec/pixel). In
the concordant $\Lambda$CDM cosmology, the conversion between
intrinsic and angular sizes is well approximated by the equation
\begin{equation}\label{Dtheta}  
\left[\frac{D}{h^{-1}_{70}\,{\mbox{kpc}}}\right]
\approx 21\frac{z(1+0.8z)}{(1+z)^2}
\left[\frac{\theta}{\mbox{arcsec}}\right]\;,  
\end{equation}
so for the adopted $\zo$ the angular pixel scale corresponds to a
physical scale of $\sim 0.5$ kpc.

The procedure followed to infer the projected density in each pixel of
the images is described in Appendix~\ref{deneval}. As explained there,
to build the continuous group images we have applied a fixed kernel
smoothing scale $h$ equal to ten times the softening length adopted
for the luminous bodies. A zero mean white sky noise component, with a
1-sigma deviation equivalent to a surface brightness of about $27$ mag
arcsec$^{-2}$, has been added to the images to allow SExtractor to
perform the photometry\footnote{The very small noise level in our mock
  images, well below the typical SDSS $r$-band sky, is intended to
  facilitate the detection of intragroup light (see
  Sec.~\ref{grp-bgg}).}. The maps have been also smoothed by a
Gaussian point spread function (PSF) with a full-width at half-maximum
of $1.4$ arcsec ($\sim 1.8$ kpc at the adopted $\zo$), a value equal
to the median PSF characterizing the SDSS-II $r$-band photometric
observations. After some trial and error, we found that for our mock
images the FLUX\_ISOCOR SExtractor method provides the best and most
accurate estimates for the total mass of the BGGs, as demonstrated by
the fact that the total initial stellar masses of our 48 simulated
groups recovered from the groups' projected image show a r.m.s.\
deviation from the corresponding actual values of only $0.02$ per
cent.
  
Once the positions and masses $\ms$ of the galaxies are inferred from
the density maps, each stellar particle is assigned to the galaxy that
maximizes the ratio $f_{ij} = M_{\star i}/R^2_{\star i,j}$, being
$R_{\star i,j}$ the projected distance between galaxy $i$ and particle
$j$. It is from the sets of luminous particles belonging to the BGGs
that we calculate the global properties of these objects. The $\Reff$
is calculated as the median of the ranked distribution of radial
distances of the particles to the centre of the galaxy, represented by
its density peak, $\SBe$ is then just the luminous mass contained
inside $\Reff$ divided by $\pi\Reff^2$, while $\sige$ is
inferred\footnote{The conclusions of this work are not affected if we
  replace the effective velocity dispersion, as measured in integral
  field spectroscopic surveys, by a central estimate, say within
  $1/8\;\Reff$.} by averaging the perpendicular component of the
velocities of the particles within $\Reff$. We have also included a
selection bias towards $\sige\gtrsim 100$ \kms\ to guarantee that we
deal only with truly massive central remnants
\citep{Mut14,BFvD15}. Then we proceeded to calculate the three
principal axes (within one $\Reff$) of the central galaxies, which
have been used to estimate the ellipticities that provide the
corrections necessary to circularize the galactocentric distances of
all the luminous particles. The new particle distances have been then
sorted to extract the integrated stellar mass profiles of the BGGs by
adding the masses of the luminous bodies contained inside a given
radius. Due to the high number of particles that accumulate near the
centre of the galaxies, we find that it is more efficient to build
these profiles applying a logarithmic binning in $R$ and accounting
for the error estimates for the mass in each bin. Finally, we have
obtained a first estimate of the global shape of the stellar mass
distributions of our mock BGGs using the median value of the S\'ersic
indices $n$ that result from applying non linear least-squares fits of
an integrated S\'ersic profile to the projections of their luminous
component onto the three Cartesian planes. In this latter calculation,
the most stable results are obtained when we exclude from the fits the
central kpc and extend the profiles until a density equal to three
times the minimum density associated with a single particle is
reached.

To compare our central merger remnants with observations mass is
converted to luminosity by considering that there are no population
effects, meaning that we assume a fixed stellar mass-to-light ratio
$\Upsilon$ for all galaxies. Since the Two-Micron All-Sky Survey
(2MASS) $K_{\rm s}$-band luminosity -- a "short" $K$ filter with a
superior suppression of thermal terrestrial emission than the original
-- is known for being a faithful tracer of the old stellar mass
content of galaxies and rather insensitive to both extinction and star
formation effects \citep[e.g.][]{KC98}, we take, after dropping the
subscript 's', $\Upsilon_K\equiv \ms/L_K = 1\;\msun/\lsun$, in line
with values reported in the literature
\citep[e.g.][]{Ell08,LS09,LaB10,McGS14}. The absolute $K$-band
magnitude measure of our remnants is then inferred from the ratio
between the total stellar mass within the model image and $\Upsilon_K$
after taking into account that ${\cal M}_{\odot,K}\simeq 3.3$
\citep{Bel03}.

\begin{figure}
\centering
\includegraphics[width=\linewidth,angle=0]{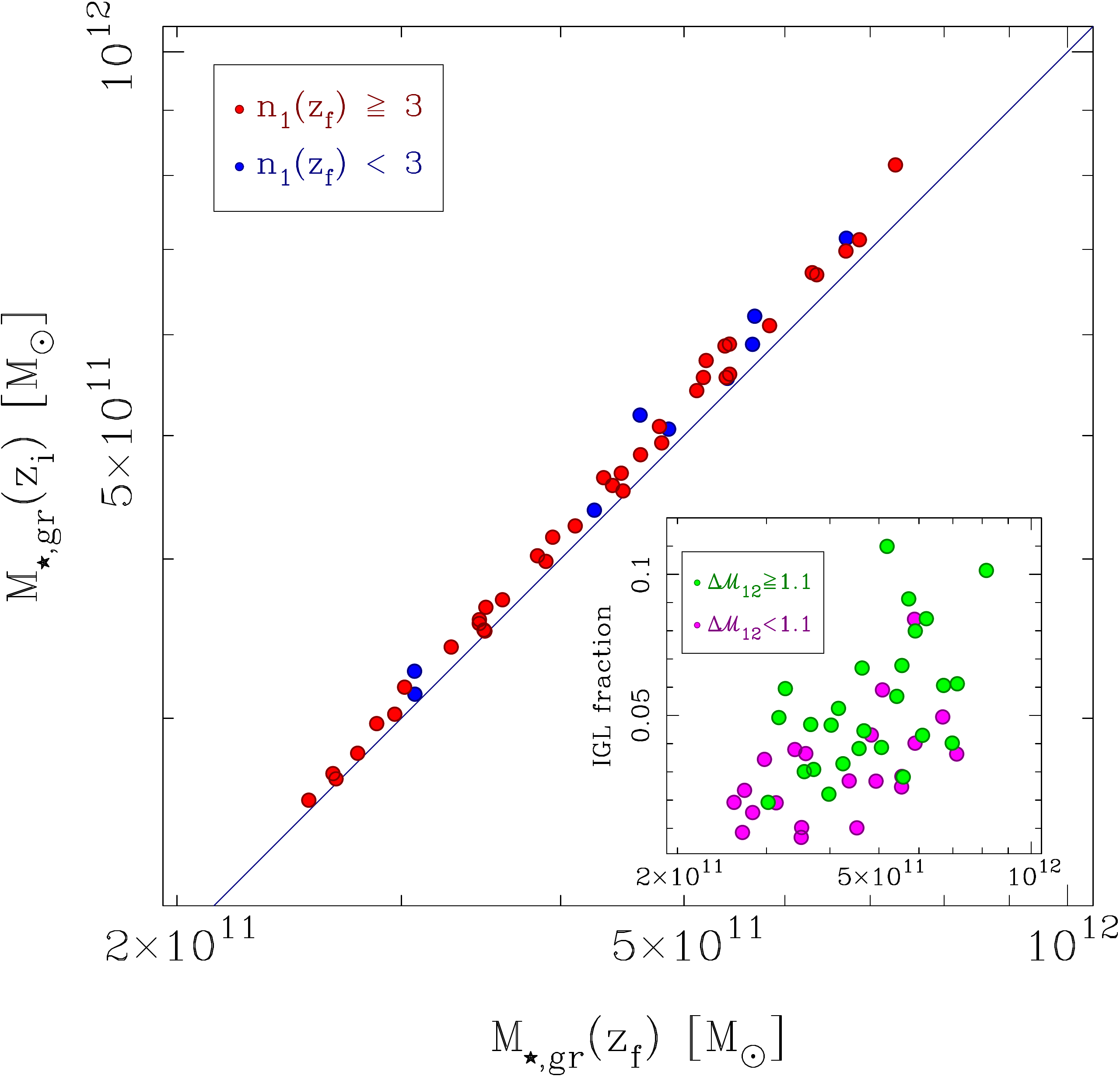} 
\caption{\small Comparison between the total mass associated with
  galaxies in our groups at $\zi$ and at $\zf$ assigned by
  SExtractor. Points filled in blue identify groups with a final
  central remnant showing a late-type-like light distribution, while
  those filled in red are used for groups having a spheroidal BGG. The
  inset allows us to see better that the difference between both
  measurements, which provides a crude estimate of the mass of the
  diffuse IGL component, tends to increase with the stellar mass of
  the group. Points in the inset are coloured according to the the
  final magnitude gap between the first- and second-ranked galaxies,
  $\Delta\magn_{12}$ (see Sec.~\ref{m12}). The adopted divider is
  close to the median of the distribution of this
  variable.}\label{fig_4}
\end{figure}

\begin{figure*}
\begin{minipage}{175mm}
\centering
\includegraphics[width=\linewidth,angle=0]{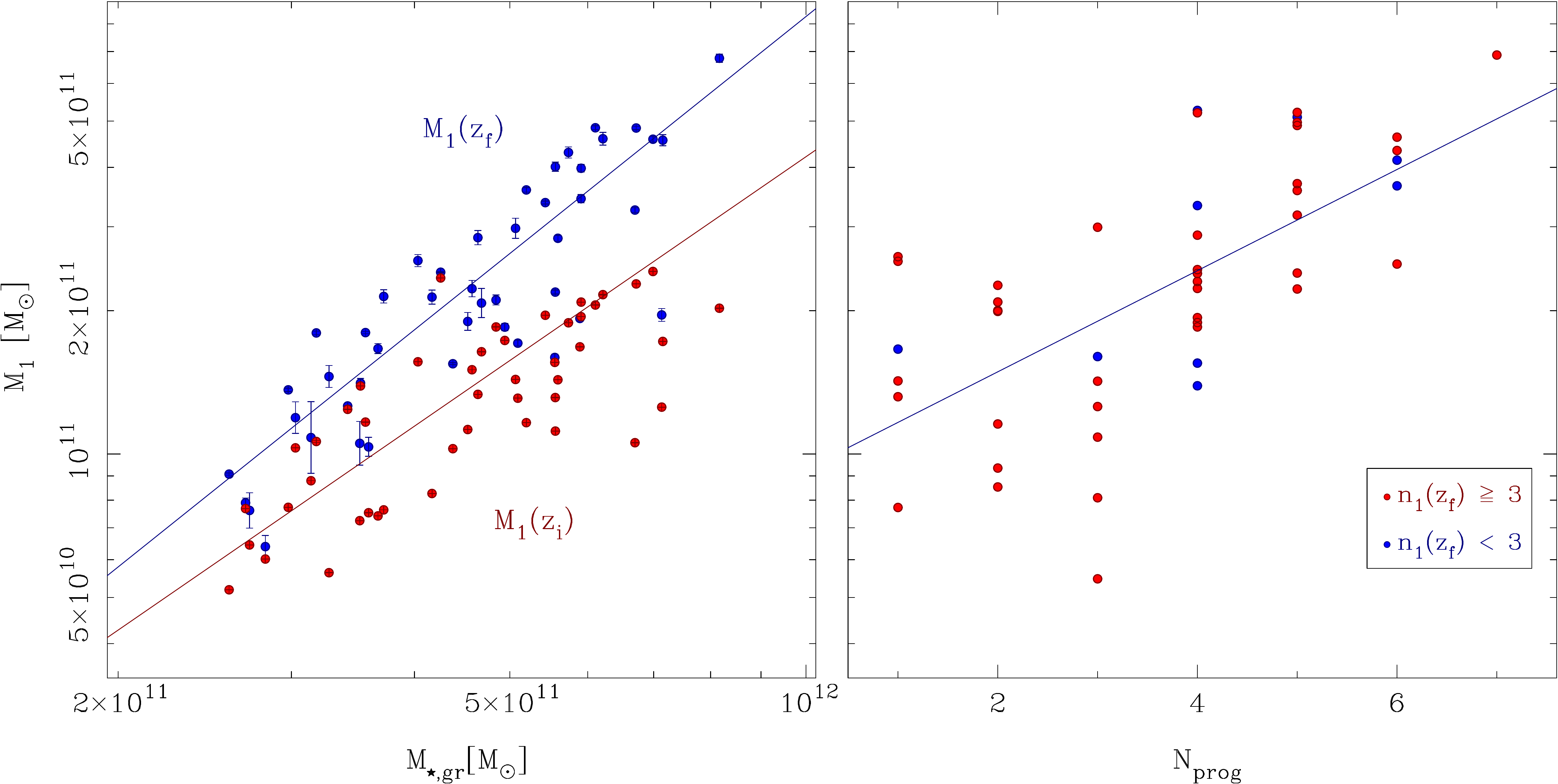} 
\caption{\small \emph{Left:} Relation between the stellar mass of the first-ranked galaxy, $M_1$, and the total stellar mass content of its parent group, $\msg$, at the beginning ($\zi$; red dots) and at the end ($\zf$; blue dots) of the simulations. Data points and associated errors have been calculated from projections of the first and final snapshots along the three Cartesian axes. \emph{Right:} Relation between the stellar mass of the first-ranked galaxy at the end of the simulations and the number of progenitors, $N_{\rm prog}$. Symbol colours are as in the main panel of Fig.~\ref{fig_4}. The straight lines used in both panels to highlight the mean trends correspond to least-squares fits to the data based on the orthogonal offsets.}\label{fig_5}
\end{minipage}
\end{figure*}

\begin{figure}
\centering
\includegraphics[width=\linewidth,angle=0]{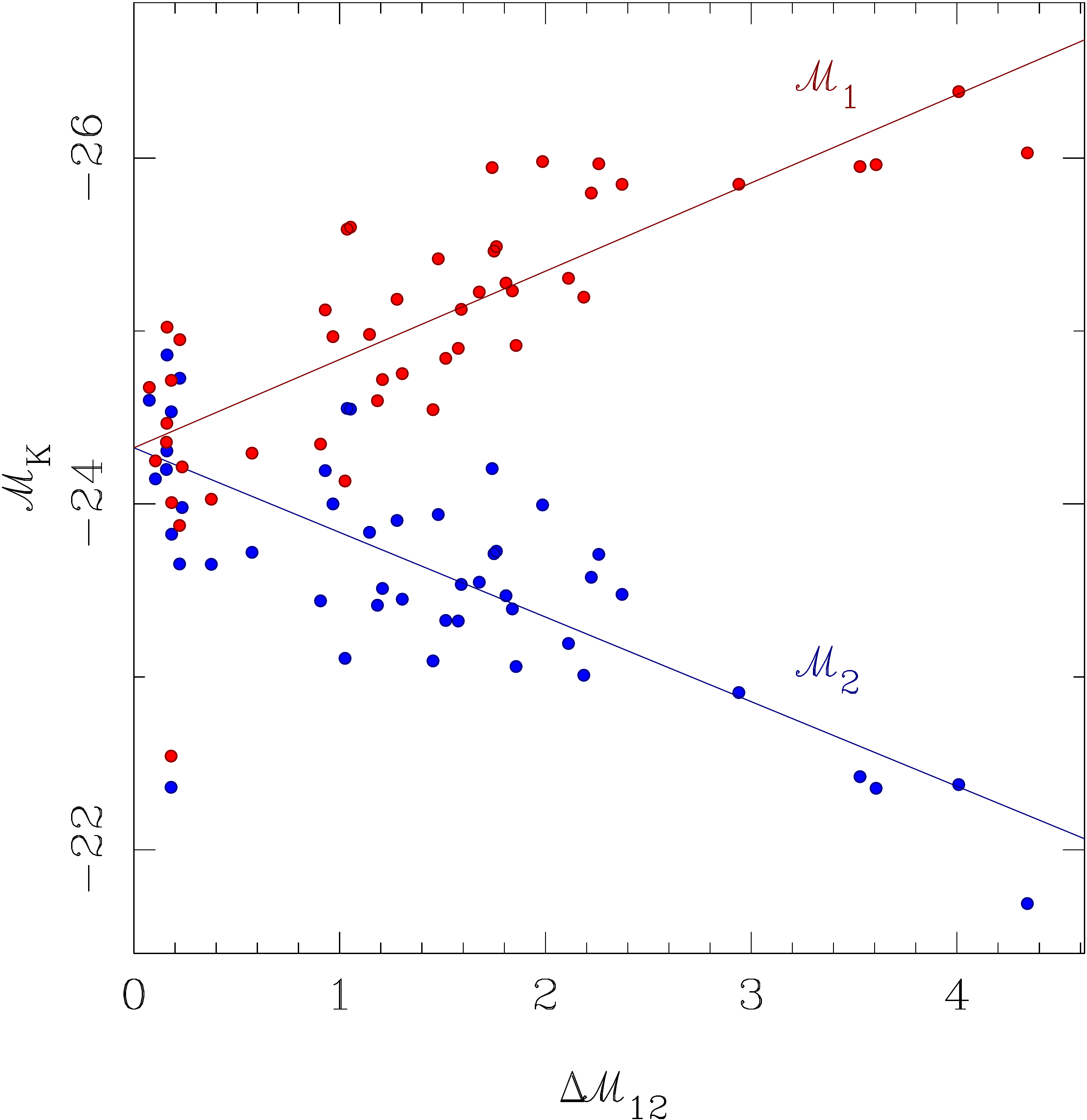} 
\caption{\small Stellar masses in ($K$-band) magnitude units of the first-, $\magn_1$ (red dots), and second-, $\magn_2$ (blue dots), ranked galaxies, created in each one of our simulated groups as a function of the difference between them, $\Delta\magn_{12}\equiv \magn_2-\magn_1$. The straight-line fits reveal that the two most luminous galaxies in our isolated forming groups follow relations of very similar strength but opposite direction.}\label{fig_6}
\end{figure} 

\begin{figure}
\centering
\includegraphics[width=\linewidth,angle=0]{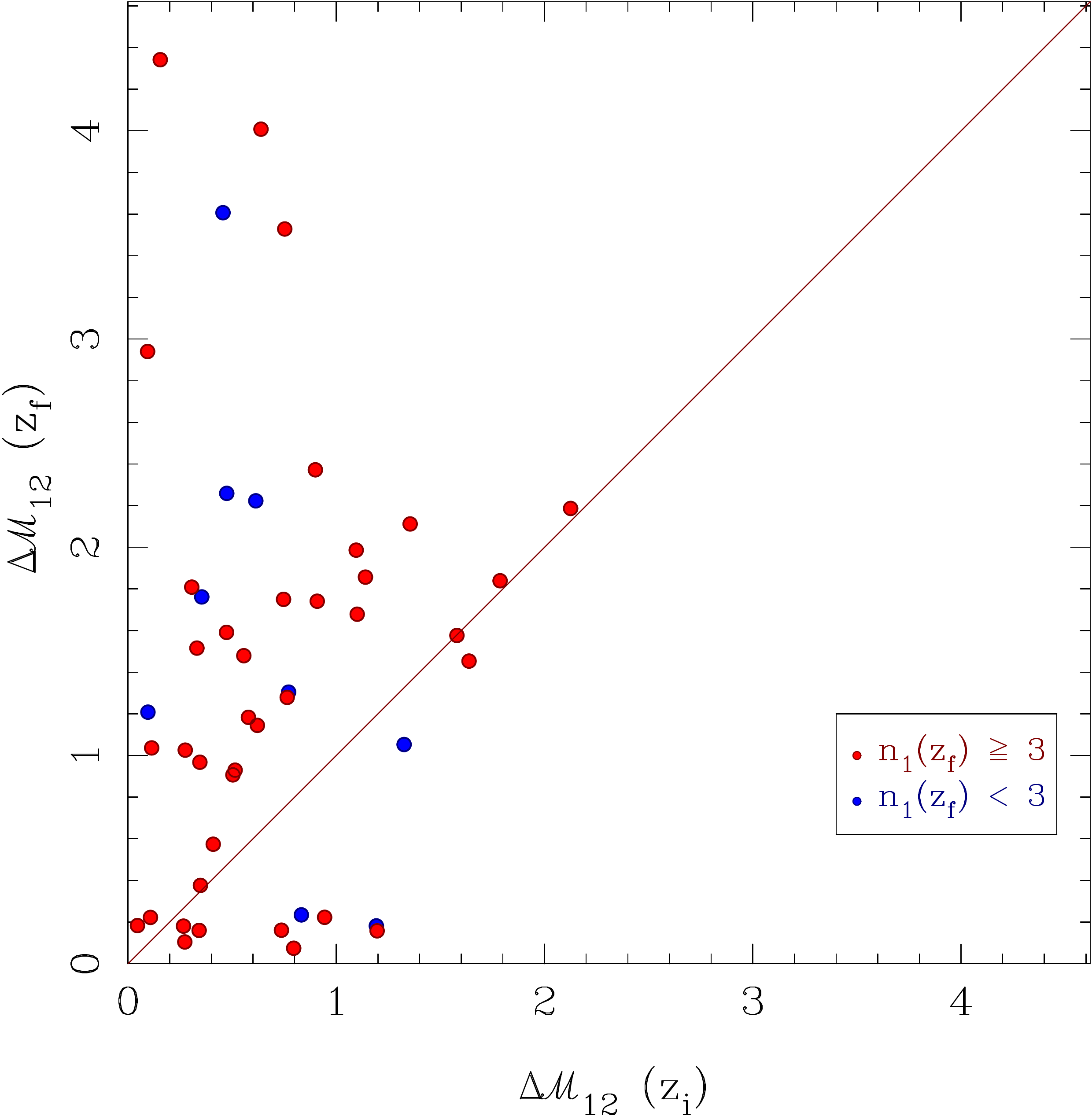} 
\caption{\small Comparison between the size of the magnitude gap,
  $\Delta\magn_{12}$, at the beginning ($\zi$) and at the end ($\zf$)
  of the simulations. Points placed above the diagonal line (towards
  y-axis) represent those cases where $\magn_1$ grows faster than
  $\magn_2$, leading to an increase in the size of the magnitude
  gap. Conversely, points placed below the diagonal line (towards
  x-axis) represent the cases in which the size of the gap gets
  smaller. Symbol colours are as in the main panel of
  Fig.~\ref{fig_4}.}\label{fig_7}
\end{figure}

\section{Optical properties of our simulated groups and BGGs}\label{optical}

In this section we aim at characterizing the importance of
collisionless gravitational dynamics during group formation by looking
in detail at the imprint it has left in the optical properties of our
systems at $\zf$. As stated in the previous section, throughout the
remaining of the paper stellar mass can be considered a proxy for
$K$-band luminosity.

\subsection{Relationship between BGG and host group properties}\label{grp-bgg}

It is evident from Fig.~\ref{fig_4} that the fraction of stars
assigned by SExtractor to individual galaxies at the end of our
simulations anti-correlates with the total stellar mass fraction of
the groups. The difference, which increases with increasing stellar
mass, barely exceeds $10\%$ in the 48 experiments analysed. Certainly,
the absence of low-mass disc galaxies in our experiments suggests that
we could be obtaining a lower limit of the amount of diffuse
intragroup light (IGL) generated during the gravitational collapse of
small galaxy groups. Besides, our measurement hardly qualifies as
slightly more than a very crude estimate of this quantity. Physically
meaningful measurements of the amount of IGL require discussing in
detail the systematics of galaxy masking and the methodology used in
the identification of this component, which is beyond the scope of
this study. In any case, let us stress the fact that obtaining such a
relatively small amount of diffuse light in such an aggressive
environment for galaxies is not new, but is a good match with the
predictions inferred from exploratory studies on the origin of the IGL
done earlier by one of us using more adequate estimators of this
component on a smaller set of simulated poor groups
\citep{Dar13}. These results suggest that explaining the abnormally
large diffuse light fractions of up to $\sim 50$ per cent of the total
light observed in a few of the compact groups catalogued by Hickson
\citep{Whi03,DRMO05} would prove problematic for pure dry merging, and
thus establish a link with the need to take into account intragroup
star formation when studying the IGL \citep[see, e.g.][]{Puc10}, if
not always, at least for these extraordinary cases.

The stellar mass (NIR luminosity) of the biggest galaxy that forms
during this process, $M_1$, also shows a tight positive correlation
with the total stellar light in the group, $\msg$, albeit with a
somewhat less marked dependence. The measured logarithmic slope of
$1.7\pm 0.1$ of the mean trend (upper curve in Fig.~\ref{fig_5}, left
panel) suggests that in isolated forming groups, of similar richness
and/or total mass, BGGs are not only brighter but tend to become more
important in the overall light as the total stellar mass content of
the groups increases. This last relationship is not merely a
reflection of the initial conditions of the simulations -- the fact
that all our groups have the same total mass and initial richness
tends to link the magnitude of their largest members at $z=3$ with the
total luminous mass -- since, as the comparison of the upper and lower
curves in the left panel of Fig.~\ref{fig_5} reveals, group formation
accentuates the dependence between both variables and their
correlation strength. Indeed, as shown in the right panel of
  Fig~\ref{fig_5}), $M_1(\zf)$, much like the final magnitude gap
between the first- and second-ranked galaxies, $\Delta\magn_{12}$ (see
next Section), displays a significant and strong positive correlation
with the number of progenitors. 

Correlations between the luminosity of the first-ranked galaxy and the
global properties of its host system have been previously noted in
observational studies \citep[e.g.][]{Han09}, and are a natural
consequence of the hierarchical evolution of structure. The difference
with respect to the results reported here is that in real galaxy
systems, first-ranked galaxies are observed to contribute a smaller
fraction of the total light with increasing mass/richness of the
hosts. This opposite behaviour, however, can be easily reconciled with
our findings considering that we are dealing with fully isolated small
forming groups. In non-isolated virialized galaxy systems any early
growth experimented by their central objects becomes later more than
compensated by the acquisition of new galaxy light from secondary
infall, and the comparatively low rate of galaxy mergers with the BGG
as groups are merged hierarchically into larger structures
\citep{LM04}.

\begin{table*}
\begin{minipage}{136mm}
\scriptsize
\centering
\caption{Compilation of \citeauthor{TR77}'s statistics}
\label{publicat}
\begin{threeparttable}

\begin{tabular}{lccrl}
\hline \hline
Acronym & $\hat{T_1}$ & $\hat{T_2}$ & $N_{\rm sys}$ & Reference(s) \\ 
\hline
2MCG    & $0.51\pm 0.06$ & $0.70\pm 0.06$  &  78 & \citet{D-G12} \\
HCG     & $1.27\pm 0.17$ & $1.01\pm 0.10$  &  67 & \citet{Hic92} \citep[see also][]{D-G12}\\ 
UZCCG   & $1.04\pm 0.15$ & $1.13\pm 0.11$  &  49 & \citet{FK02} \citep[see also][]{D-G12}\\
LCCG    & $1.10\pm 0.27$ & $1.10\pm 0.19$  &  17 & \citet{AT00} \citep[see also][]{D-G12}\\
mG11CG  & $0.46\pm 0.02$ & $0.59\pm 0.02$  & 326 & \citet{D-G12} \\
SLRG    & $0.34\pm 0.03$ & $0.72\pm 0.06$  & 210 & \citet{LS06} \\
SLRG-cl & $0.75\pm 0.10$ & $0.85\pm 0.12$  &  60 &  \\ 
SLRG-gr & $0.27\pm 0.06$ & $0.42\pm 0.09$  &  40 &  \\
SC4     & $0.93\pm 0.01$ & $0.95\pm 0.01$  & 494 & \citet{LOM10} \\   
SC4-hL  & $0.70\pm 0.01$ & $0.96\pm 0.01$  & 124 &  \\
SC4-lL  & $0.84\pm 0.01$ & $0.94\pm 0.01$  & 370 &  \\
SRXFG   & $0.23\pm 0.06$ & $0.24\pm 0.05$  &  10 & \citet{Pro11} \\ 
SMr18   & $0.99\pm 0.04$ & $1.00\pm 0.03$  & 743 & \citet{Ber06} \\  
mLMG    & $0.64\pm 0.15$ & $1.05\pm 0.18$  &  48 & This work \\ \hline
\end{tabular}
\begin{tablenotes}
\item \hspace{-0.75mm}{\bf Notes.} Errors for $\hat{T_1}$ and $\hat{T_2}$ are standard deviations computed from bootstrap. {\bf Descriptions.} 2MCG: Complete and well-defined full-sky sample of CGs selected by stellar mass from the 2MASS extended source catalogue with radial velocity larger than 3000\ \kms\ and 4 or more concordant members; HCG: \citeauthor{Hic92}'s CGs with radial velocity larger than 3000\ \kms\ and 4 or more concordant members; UZCCG: CGs in the 3D UZC galaxy catalogue with radial velocity larger than 3000\ \kms\ and 4 or more concordant members; LCCG: CGs in the Las Campanas Redshift Survey with radial velocity larger than 3000\ \kms\ and 4 or more concordant members; mG11CG: Mock CGs with 4 or more concordant members extracted from the galaxy outputs of the \citet{Guo11} semi-analytical model run on top of the Millennium-II cosmological simulation; SLRG: Fields with luminous ($\gtrsim 3\;\lc$) red galaxies (LRGs) at $z< 0.38$ selected from the SDSS-DR1; SLRG-cl: Cluster-sized SDSS LRG fields with richness in the upper 75th percentile; SLRG-gr: Group-sized SDSS LRG fields with richness in the 25th-50th percentile; SC4: Updated version of the C4 cluster catalogue extracted from the SDSS-DR5 restricted to systems with $\sigma_{\rm los} > 200$\ \kms\ at $z=0.030$--$0.077$; SC4-hL: High-luminosity subsample ($L_{\rm tot}>L_{\rm div}=3.7\times 10^{11}\;h_{70}^{-2}\lsun$) of the SC4 clusters; SC4-lL: Low-luminosity subsample ($L_{\rm tot}\le L_{\rm div}$) of the SC4 clusters; SRXFG: High X-ray luminosity fossil groups ($L_X > 5\times 10^{41}\;h_{70}^{-2}\,\mbox{erg}\;\mbox{s}^{-1}$) with a greater than 2-mag $i$-band $\magn_{12}$ gap within half the virial radius selected from the SDSS and RJX samples; SMr18: Volume-limited spectroscopic sample of galaxy groups and clusters with $0.02 < z < 0.05$ complete down to $\magn_r-5\log h<-18$ mag identified from the SDSS-DR3; mLMG: Low-mass ($M_{\rm gr}= 10^{13}\;h^{-1}\msun$) collapsing groups created in this work from cosmologically-consistent controlled simulations.
\end{tablenotes}
\end{threeparttable}
\end{minipage}
\end{table*}

\subsection{The nature of the highest ranked galaxies}\label{m12}

After detecting the relationship between the properties of BGGs and
of their host groups, it is natural to ask if this connection extends
to other luminous (massive) group members.

We address this question in Fig.~\ref{fig_6}, where we plot the
stellar masses, expressed in ($K$-band) magnitude units, of the first-
and second-ranked galaxies identified at $\zf$ in each one of our
groups, $\magn_1$ and $\magn_2$, respectively, as a function of their
difference $\Delta\magn_{12}\equiv \magn_2-\magn_1$. Examination of
the plot reveals in the first place that our simulated groups
encompass quite a broad range of $\Delta\magn_{12}$ values, from
nearly zero up to more than 4 mag. This result shows that, even though
our groups contain a fixed number of initial members drawn from the
same mass function and start at the same evolutionary stage, the
randomness of the initial positions of their member galaxies in phase
space, and of the sampling of morphologies and masses, combine with
their highly non-linear evolution to guarantee that each particular
realization of the initial conditions leads to a unique and varied
assembly path. The most striking result shown by this plot is,
however, the fact that $\magn_1$ and $\magn_2$ follow scalings with
$\Delta\magn_{12}$ of very similar strength but of opposite sign:
positive for the former and negative for the
latter.\footnote{Incidentally, note that the current set of axes of
  Fig.~\ref{fig_6} means that we are effectively plotting $y-x$ vs
  $x$. Given that the slope of the red line is nearly $0.5$, the
  expected slope of the blue line should be around $0.5-1=-0.5$ as
  observed. We thank the referee for pointing this out in his report.}
Such an outcome indicates that first-ranked group objects tend
generally to grow in brightness at the expense of their lesser
companions during the build-up of their host galaxy aggregations, so
that most of the magnitude gaps detected at the end of our simulations
arise from the brightening of the BGGs and the dimming of the second
brightest galaxies in about equal measure. However, as illustrated in
Figure~\ref{fig_7}, this scenario cannot be fully extrapolated to all
the groups for, among those with a final $\Delta\magn_{12}\lesssim 1$
mag), in nearly half of cases $\magn_2$ actually grows faster than
$\magn_1$, leading to a reduction in the magnitude gap size. This last
Figure also shows that in our simulations the initial breadth of the
gaps does not predispose them to achieve a particular final size owing
to the total lack of correlation between the initial positions of the
galaxies and their luminosities. To all this it must be added the fact
that the tight links between the stellar mass of the first-ranked
object and both $\Delta\magn_{12}$ and $\msg$ (see
Sec.~\ref{grp-bgg}), suggest that a similarly strong relationship
should exist between the magnitude gap and the global stellar mass
fraction of the parent halo, as it can be inferred from the spatial
segregation shown by the data points in the inset of
Figure~\ref{fig_4} when we use different colours to divide $\msg(\zf)$
according to gap size.

\begin{figure*}
\begin{minipage}{175mm}
\centering
\input{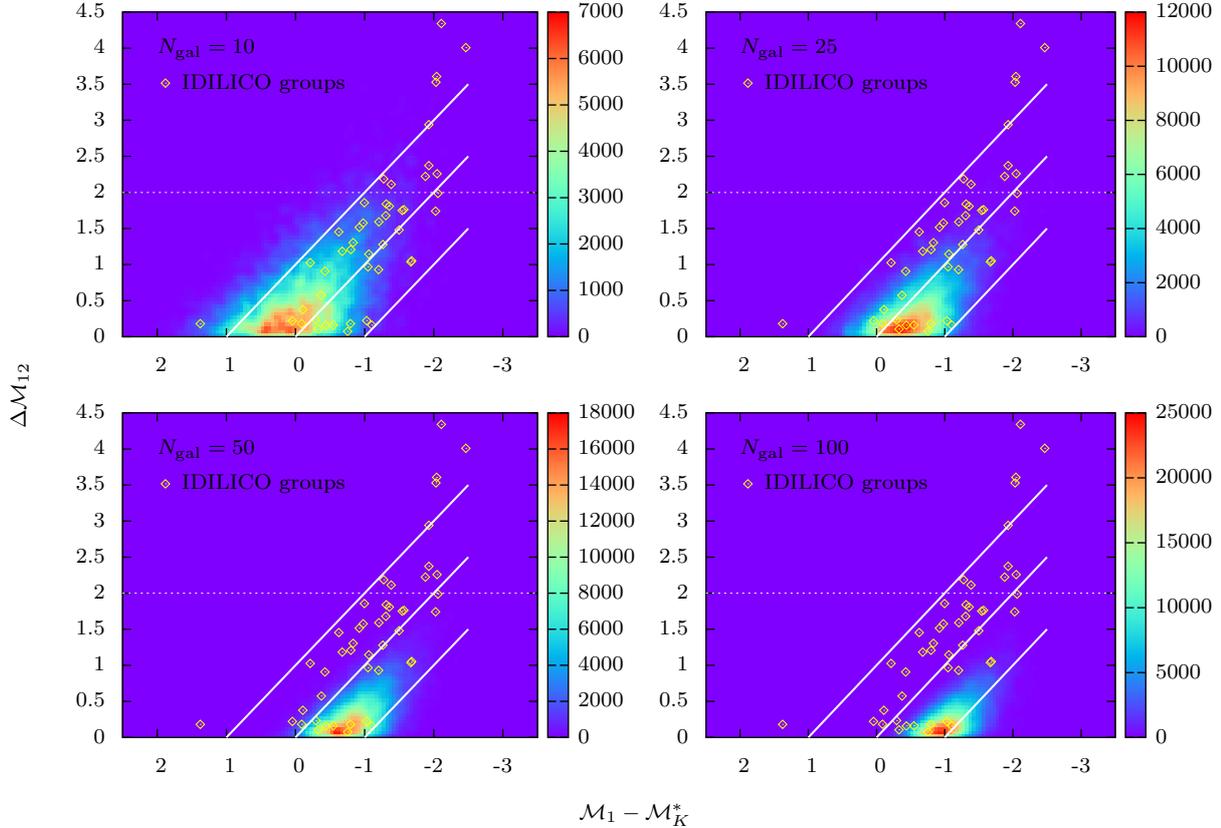}
\caption{\small The difference $\Delta\magn_{12}$ versus the absolute
  $K$-band magnitude of the first-ranked galaxy, $\magn_1$, relative
  to the characteristic luminosity of the observed \citeauthor{Sch76}
  $K$-band LF for the groups studied in this work. We show values
  inferred from the median stellar mass using
  $\Upsilon_K=1\;\msun/\lsun$, and $\magn_{K}^*=-23.95$ from
  \citet*{SLC09} (open yellow diamonds). The coloured 2D histograms on
  the background of each of the four panels depict the joint
  distributions of values of $\magn_1-\magn^*_K$ and
  $\Delta\magn_{12}$ expected from the null hypothesis that the
  luminosity of the first-ranked object in a galaxy aggregation of
  richness $N_{\rm gal}=10$, 25, 50, and 100, is simply an extreme
  value of a universal \citeauthor{Sch76} LF with asymptotic slope
  $\alpha=-1.0$ truncated at $L<0.05\;\lc$. All histograms are based
  on equal-size sets of $10,000$ two-dimensional data points (note the
  different scales of the colour bars depicting pixel counts). White
  diagonal lines represent the loci of constant second-ranked galaxy
  luminosities; from top to bottom $\magn_2-\magn^*_K=1.0$, $0.0$, and
  $-1.0$~mag. The horizontal white dotted line represents the typical
  threshold in $\Delta\magn_{12}$ used in the literature to define
  fossil groups.}\label{fig_8}
\end{minipage}
\end{figure*}

\newpage

\begin{figure}
\centering
\includegraphics[width=\linewidth,angle=0]{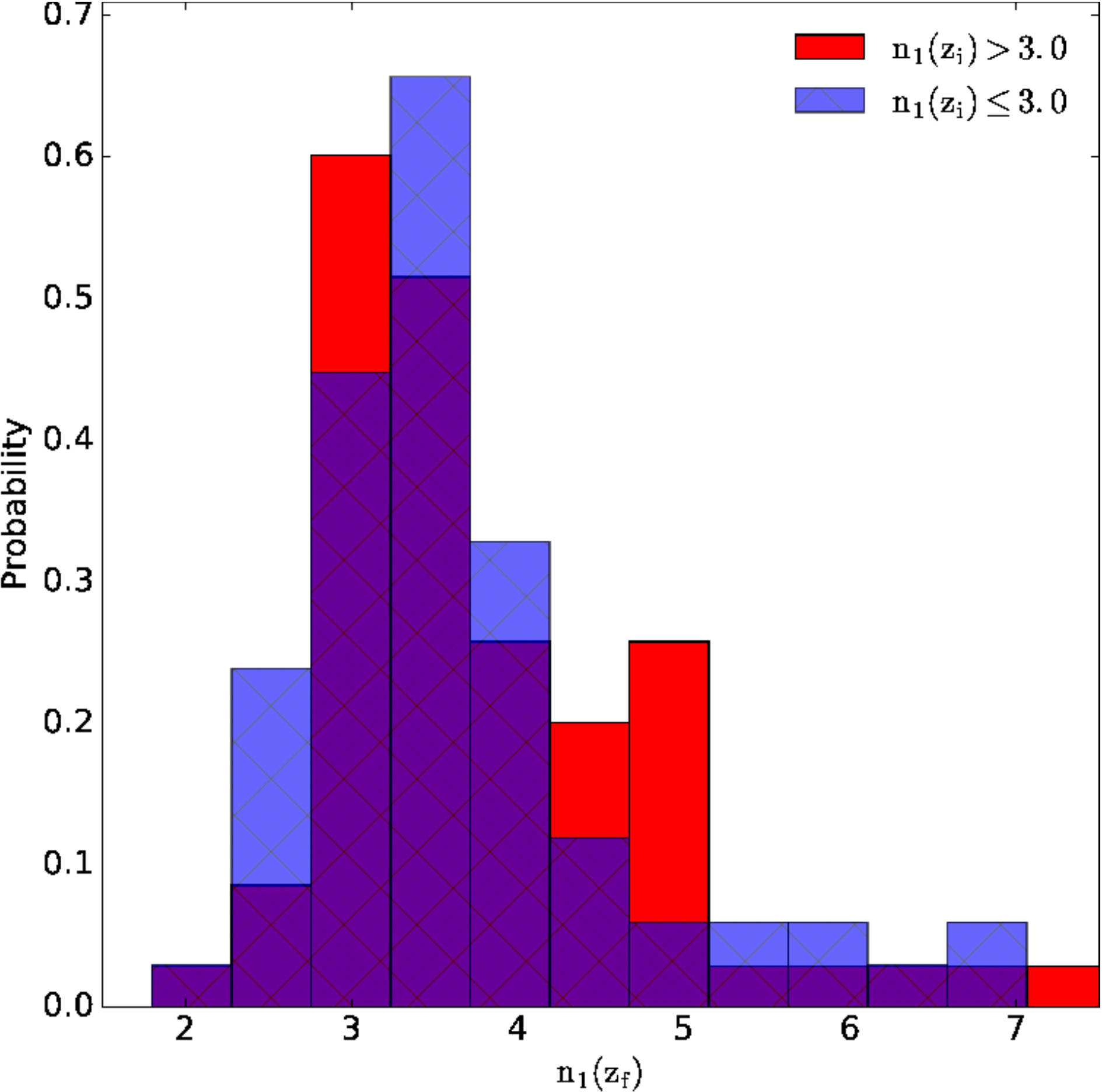} 
\caption{\small Distribution of the final S\'ersic indices of our
  simulated BGGs, $n_1(\zf)$, as a function of the S\'ersic index of
  the first-ranked member of their groups on input, $n_1(\zi)$. The
  value adopted as divider separates those galaxies with a disk and
  those with a spheroid as the most massive progenitor. Note that we
  are showing fractional distributions, so the histograms are
  normalized to the same (arbitrary) area.}\label{fig_9}
\end{figure} 

\begin{figure}
\centering
\includegraphics[width=\linewidth,angle=0]{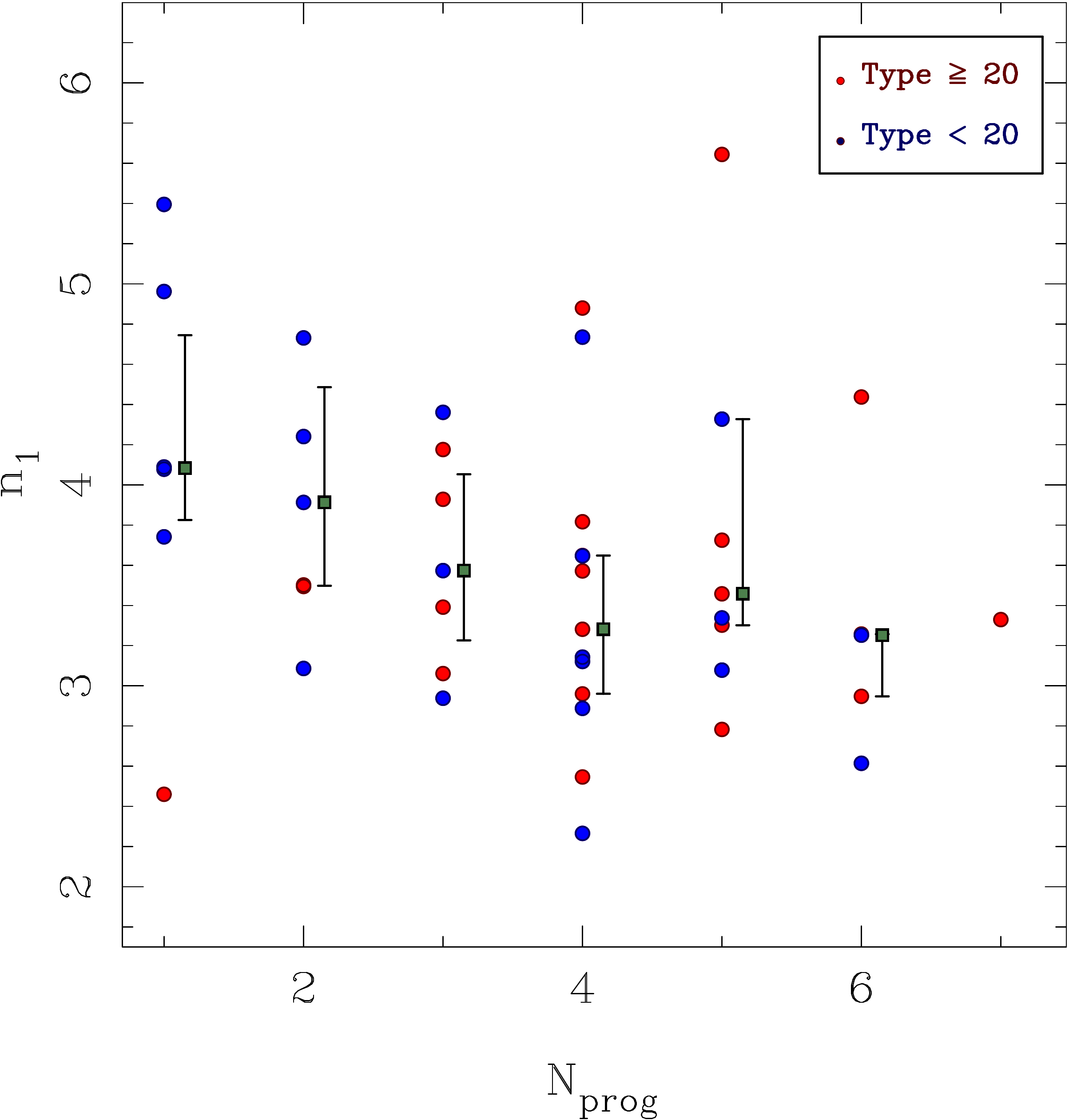} 
\caption{\small S\'ersic index of the final first-ranked members of
  the groups, $n_1$, as a function of the number of relevant
  progenitors $N_{\rm prog}$ (i.e.\ galaxies which contribute at least
  $5\%$ to the final mass of the BGG) that have participated in their
  assembly. Points are coloured according to the variable $Type$,
  which accounts for the percentage (from 0 to 100) of the BGG stellar
  mass provided by spheroidal progenitors (see text). The adopted
  divider splits the data into two subsets of similar size separating
  BGGs arising mostly from discs from the rest. Large green squares
  with vertical error bars show the median values and upper and lower
  quartiles of $n_1$ for the different values of $N_{\rm prog}$. They
  have been offset to the right on the x-axis for
  clarity.}\label{fig_10}
\end{figure}

\begin{figure*}
\begin{minipage}{175mm}
\centering
\includegraphics[width=\linewidth,angle=0]{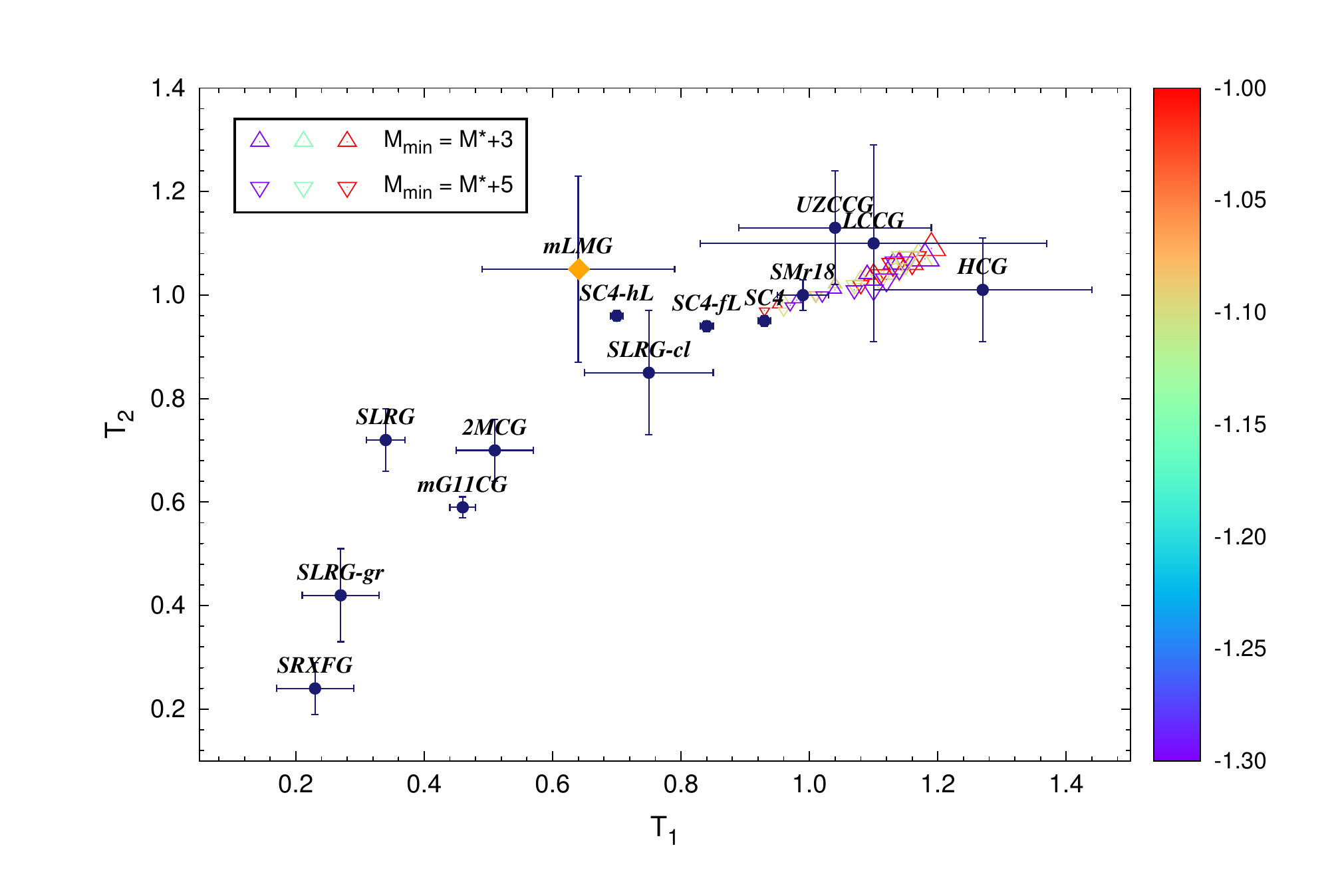} 
\vspace{-30pt}
\caption{\small Graphical representation of the estimates of the \citeauthor{TR77} statistics listed in Table~\ref{publicat}. The pair of values $(\hat{T_1},\hat{T_2})$ inferred from our collection of simulated poor groups at $z\sim 0$ is highlighted with a large filled orange square. Error bars depict $1\,\sigma$ uncertainties from bootstrap. The open triangles in the background show the expectation values of the statistical hypothesis inferred from a \citeauthor{Sch76} LF. Triangle colours represent different values of $\alpha$ (see the right bar), while their orientations and sizes inform, respectively, about the value of the faint-end cutoff (see the inset) and group richness: from 5 (smaller) to 500 (larger).}\label{fig_11}
\end{minipage}
\end{figure*}

Let us now be a bit more specific regarding the merger history of the
highest-ranked galaxies. Within our mock forming groups, first-ranked
galaxies experience between 0 and 6 minimally relevant mergers
(i.e.\ involving galaxies which contribute at least $5\%$ to the final
mass of $\magn_1$), the mean number of mergers being 2.64 and the mode
(most likely value) three, meaning that our BGGs have typically at
least 4 progenitors above a certain size (even so, we find 6 BGGs with
a single main progenitor). In 47 out of our 48 groups the first-ranked
member at the beginning of the simulations is also a progenitor of the
final BGG, while in 45 groups all the relevant progenitors of the BGG
are among the ten biggest input galaxies of each system, which
explains why the final stellar mass of the BGGs shows a strong
positive linear relationship with the number of relevant
progenitors. In contrast, we find that final second-ranked galaxies
experience between 0 and 4 minimally relevant mergers, with a mean of
1.08, and a modal value of zero, i.e.\ a single main progenitor,
sampled in 18 groups (in half of these, the progenitor was a third- or
lower-ranked input galaxy). Besides, only in half of our groups (24)
the final second-brightest galaxy contains its counterpart at the
beginning of the simulations, while in other 23 cases the two
brightest input galaxies eventually end up as part of the final
BGG. Our simulations therefore suggest that the properties of central
objects in dynamically evolved galaxy systems are determined mainly by
the assembly of the parent structure \citep{Vul14}, a process that we
also find is unique to this class of galaxies and does not appear to
significantly affect lesser system members. This implies, in good
agreement with the findings by \citet{LOM10}, \citet{D-G12}, and
\citet{She14} among others, that BGGs, unlike their second-ranked
companions, cannot be in general the statistical extreme, in terms of
their luminosity, of the group/cluster galaxy population.

The information provided by Fig.~\ref{fig_6} is completed with
Fig.~\ref{fig_8}, which shows the magnitude gap $\Delta\magn_{12}$ in
our simulated groups against the magnitude of their first-ranked
galaxy measured in units of the observed characteristic scale of the
Schechter LF of $K$-band galaxies (open yellow diamonds). This plot
contains four panels, each showing the expected joint distribution of
values of $\Delta\magn_{12}$ and $\magn_1-\magn^*$ for groups of a
given richness, $N_{\rm gal}$, whose galaxies obey a Schechter LF with
shape parameters close to those used to sample the masses of our
simulated galaxies. An examination of Fig.~\ref{fig_8} reveals that
our first-ranked galaxies tend to be abnormally bright,
$\magn_1\lesssim\magn^*$~mag, for groups of richness $N_{\rm gal}\leq
25$, even in systems with a moderate bright-end gap,
$\Delta\magn_{12}\leq 1$~mag. At the other end of the gap spectrum, we
find that groups with $\Delta\magn_{12}>2$~mag, are characterized by a
small spread in $\magn_1$, with typical $\magn_1$ values around
$\magn^*-2.0$~mag, and second-ranked galaxies quite faint, with 6 out
of 10 fossil-like groups having $\magn_2\gtrsim\magn^*+1$~mag. These
results altogether confirm that, in a substantial number of cases, the
luminosities of our BGGs deviate significantly from the statistical
expectations, implying that these galaxies, and just them, are a
completely different population from other group members. Also note
the broad dynamic ranges covered by the two variables: $\sim 2.5$
magnitudes for $\magn_1$ and, especially, $\sim 4.5$ magnitudes for
$\magn_{12}$. They show quantitatively that, as discussed above
(Sec.~\ref{simul_set-up}), the narrow ranges of group masses and
memberships explored by our simulations have not proven to be an
obstacle in producing a wide variety of final group configurations
(see also the discussion regarding the statistics $T_1$ and $T_2$
below).

First- and second-ranked objects are also at variance regarding their
radial distribution of starlight, which we have parameterized using
the S\'ersic shape index, $n$, derived from fits to the integrated
light profile. The differences, however, are to be found below $n=4$,
as the cumulative distribution functions of $n$ for both galaxy
subtypes look pretty much the same above this limit, which accounts
for $\sim 25\%$ of all the measurements. First-ranked galaxies show a
relatively narrow distribution of values of $n$, with about $70\%$ of
them having light profiles reasonably well described by the
\citeauthor{dVac48}' \citeyear{dVac48} $R^{1/4}$ surface-density law,
a special case of the S\'ersic's law with index $n=4$. In contrast,
$\sim 50\%$ of the second-ranked galaxies have $n\lesssim 3$. On the
other hand, as we show in Figure~\ref{fig_9}, the structure of the
precursor galaxies does not seem to make an appreciable impact on the
light distribution of the central remnants \citep[but see e.g.\ Fig.\
6 in][]{TDY13}. This Figure depicts the fractional distribution of
final S\'ersic indices for our BGGs, $n_1(\zf)$, split into objects
having a most massive input progenitor with S\'ersic index
$n_1(\zi)\geq 3$ (spheroid) and those in which $n_1(\zi)<3$ (disc).
The high degree of similarity shown by the blue and red histograms
(corroborated by a two-sample Kolmogorov--Smirnov test) indicates that
the final shapes of central galaxies do not correlate with the initial
morphologies of their main progenitors. Another factor that might
possibly be closely related to the structure of the central remnants
is the number of mergers they experience during their assembly. In
Figure~\ref{fig_10} we depict $n_1(\zf)$ against the number of
relevant progenitors, $N_{\rm prog}$. This plot shows that,
notwithstanding the large dispersion in the final values of $n_1$
(also plain in Fig.~\ref{fig_9}), the central galaxies emerging from
our simulations have typical inner structures characterized by
S\'ersic indices $\sim 3.5$--4 whatever the number of merged
progenitors, even in those cases in which $N_{\rm prog}$ is reduced to
the minimum\footnote{Note that the absence of mergers does not
  preclude galaxies for becoming dynamically hot through, e.g.,
  recurrent collisions with small satellites or strong tidal
  interactions with neighbours.}. It is worth noticing that both the
spread of S\'ersic index values and its cumulative distribution above
$3.0$ match very closely the data from the BGGs present in the fossil
groups studied by \citet{MenA12}.

Regarding the lack of correlation between $n_1(\zf)$ and other
variables, such as $N_{\rm prog}$, it must be noted that the evolution
of a dense system with about two dozens of galaxies that is collapsing
under its own gravity is necessarily very complex. On this stage,
group members not only may experience multiple mergers, but also, and
independently, frequent collisions with neighbours that will produce
all kinds of tidal disturbances. Besides, the design of our runs does
not guarantee that all the resultant mock BGGs end up in a similar
dynamical state. All this makes the behaviour of the final S\'ersic
index of our BGGs not as predictable as it might seem at first glance,
as perfectly illustrates Figure~\ref{fig_10}. This plot shows that the
precursors of the single-progenitor BGGs with the five highest values
of $n_1(\zf)$, all above $\sim 3.7$, are all actually disc galaxies,
while the BGG with $n_1(\zf)\sim 2.5$ arises from a spheroid. The
review of the fitting procedure, the random behaviour of the
residuals, and the consistency of fit and SExtractor magnitudes, have
convinced us that this is not a spurious result due to a bug in our
fitting code, or to the small noise level assumed in our mock images,
or to incorrect masking. At the same time, the study of the evolution
of these objects shows that they are all relatively isolated large
galaxies surrounded by multi-generation shell systems formed via
recurrent collisions with a sizable number ($4$--$6$) of small
satellites -- it is precisely the low mass of the satellites that
prevents them from merging with the central object for the duration of
the runs. The absence of relevant mergers, however, has not prevented
the main galaxies from becoming dynamically hot. In the case of the
low-$n$ BGG, we have noted that the multiple passages of the
satellites across the body of the main galaxy take place mostly along
the same plane, which could favour phase mixing in the central regions
and, therefore, the formation of a remnant with a less concentrated
surface brightness distribution. Even so, the possibility that fits by
a single S\'ersic profile may not be the best method to characterize
the stellar structure of such strongly perturbed galaxies should not
be ruled out either. Either way, one would expect $n_1(\zf)$ to be a
reliable indicator of the overall structure for the well-mixed
multi-merger systems.

In line with the last results, nor we have found in our simulations
clear signs that the light distribution of first-ranked galaxies is in
any way related to their stellar mass, the luminous mass of the host
group or the sampling of the masses of the progenitors (see
Figs.~\ref{fig_4} (main), \ref{fig_5} (right) and \ref{fig_7}). Also
note that in Figure~\ref{fig_10} data points have been coloured
according to the mass-weighted morphology of \emph{all the relevant}
input progenitors, represented by the variable $Type$, which accounts
for the percentage of the stellar mass of the BGG originated from
spheroidal galaxies. We use a value of $Type=20$ to separate those
BGGs made mostly from discs from the rest. The absence of a clear
segregation in the distribution of colours indicates that the
morphological mix of progenitors has a negligible effect on the final
shapes of central galaxies. This has been confirmed by applying a
two-sample K-S test to the resulting probability distribution
functions (PDFs) of final S\'ersic indices.

We have seen that the effects of dynamical friction are specially
noticeable in the brightest (more massive) members of a group.
Therefore, the relative brightening of first-ranked galaxies holds
important clues about the dynamical history of their local environment
and can be used to compare different samples of galaxy associations.
\citet{TR77} devised a couple of statistics related to the
$\Delta\magn_{12}$ gap suitable for this task. These authors realized
that within the \emph{statistical scenario} the average magnitude
difference between the first- and second-ranked members of a galaxy
system must be of the same order as the size of the spread of both
this magnitude difference, $\sigma(\Delta\magn_{12})$, and the
brightest galaxy magnitude, $\sigma(\magn_1)$, provided only that the
numbers of galaxies in non-overlapping magnitude intervals behave as
independent random variables \citep{Sco57}. More specifically,
\citeauthor{TR77} demonstrated that under the hypothesis of
independent luminosity sampling any set of galaxy aggregations built
from ordinary LFs\footnote{The LF does not have to be universal or
  continuous.} must verify the following two
inequalities:\footnote{Note that $T_1$ and $T_2$ deal with magnitude
  differences of companion galaxies of the same (early) type. This
  eliminates first-order uncertainties associated with colour
  conversions, as well as redshift- and direction-dependent
  corrections, facilitating the comparison of data from heterogeneous
  sources.}
\begin{equation}\label{t1t2}
T_1\equiv\frac{\sigma(\magn_1)}{\langle \Delta\magn_{12}\rangle}\ge 1\ \ \ \ T_2=\frac{1}{\sqrt{0.677}}\frac{\sigma(\Delta\magn_{12})}{\langle \Delta\magn_{12}\rangle}\gtrsim 1\;.
\end{equation}
Galaxy merging within isolated groups should reduce the values of
$T_1$ and $T_2$ well below unity, making them inconsistent with random
sampling of luminosity functions, as indicated by the results from
both observations and cosmological simulations \citetext{see
  \citealt{D-G12} and references therein}.

The two statistics defined in eqs.~(\ref{t1t2}) are used in
Fig.~\ref{fig_11} to place our collection of simulated groups in
relation to existing observed and mock catalogues of galaxy
associations for which the values of $T_1$ and $T_2$ are either known
or easily calculable from published data. Selected data sets include
several compilations of compact groups (CGs), but also X-ray fossil
groups, clusters, and fields with luminous red galaxies. For clarity,
the main specifications of the catalogues, the number of galaxy
systems included, $N_{\rm sys}$, and their estimated values of $T_1$
and $T_2$ are summarized in Table~\ref{publicat}. Note that in some
cases, we also present results for subsets of these catalogues arising
from partitions based on richness or luminosity.

Since the theoretical lower limits of eqs.~(\ref{t1t2}) are
derived under very general conditions about the form of the LF and
considering galaxy aggregations with a large number of members, we
have also included in Fig.~\ref{fig_11} the expectation values for
$T_1$ and $T_2$ that result from the statistical scenario under more
realistic conditions. Specifically, we assume that the underlying LF
can be reasonably approximated by a Schechter-type functional
truncated at either $\magn \le \magn^*+3$~mag or $\magn \le
\magn^*+5$~mag, with steepness $\alpha$ ranging from $-1.0$ to $-1.3$
and normalization (richness) varying from 5 to 500 members. The
averages $\langle \hat{T_1}\rangle$ and $\langle \hat{T_2}\rangle$
that we calculate from $10,000$ Monte Carlo realizations of each
possible combination of the above constraints (coloured empty
triangles in Fig.~\ref{fig_11}) are mainly sensitive to richness,
moving from $\sim1.2$ and $1.1$, respectively, when both estimators
are inferred for the largest galaxy associations ($N_{\rm gal}=500$),
to values marginally below the estimated thresholds when they are
calculated for the poorest groups ($N_{\rm gal}=5$).

As shown by Fig.~\ref{fig_11} and Table~\ref{publicat}, the estimates
of the \citeauthor{TR77} statistics for our groups at $z\sim 0$ taken
together are incompatible with the predictions from the statistical
scenario. Still, the discrepancy essentially arises from the fact that
our simulated groups show quite a broad first-second magnitude gap
compared to the dispersion in $\magn_1$ (see also Fig.~\ref{fig_6}),
which leads to a value of $\hat{T_1}=0.64\pm 0.15$ significantly
smaller than the predictions of the statistical scenario within the
quoted $1\,\sigma$ uncertainties inferred from $10,000$ bootstraps. In
contrast, the estimator $\hat{T_2}=1.05\pm 0.18$ is consistent with
random sampling conditions because of the large dispersion in
the values of $\Delta\magn_{12}$ that characterizes our groups. This
difference in the behaviour of the \citeauthor{TR77} statistics, in
which the first one points to the non-statistical origin of the data
set but not the second, has been observed in complete and well-defined
samples of groups/clusters of galaxies by \citet{LOM10} and
\citet[][see their Fig.~13]{She14}, suggesting that $T_1$ is perhaps a
more sensitive estimator of the true nature of BGGs than $T_2$. Taken
together, our collection of mock isolated forming galaxy aggregations
represents an intermediate situation between several non-BGG dominated
compact group and cluster catalogues, including the spectroscopic
followup by \citet{Hic92} of the well-known \citeauthor{Hic82}'s
compact groups \citep[][hereafter, HCGs]{Hic82} -- whose full
compatibility with the statistical hypothesis likely emanates from the
use of selection algorithms biased against the inclusion of very
dominant galaxies --, and the extremely merger-driven fossil groups
selected by \citet{Pro11}, which are characterized by significantly
small values of $\hat{T_i}$ (see Fig.~\ref{fig_11}). In very good
agreement with these latter observations, our 10 groups verifying the
condition $\Delta\magn_{12}>2$ mag yield $\hat{T_1}=0.12\pm 0.03$ and
$\hat{T_2}=0.35\pm 0.05$.

\subsection{Evolution of the galaxy mass/luminosity function}\label{starMF}

\begin{figure}
\centering
\includegraphics[width=\linewidth,angle=0]{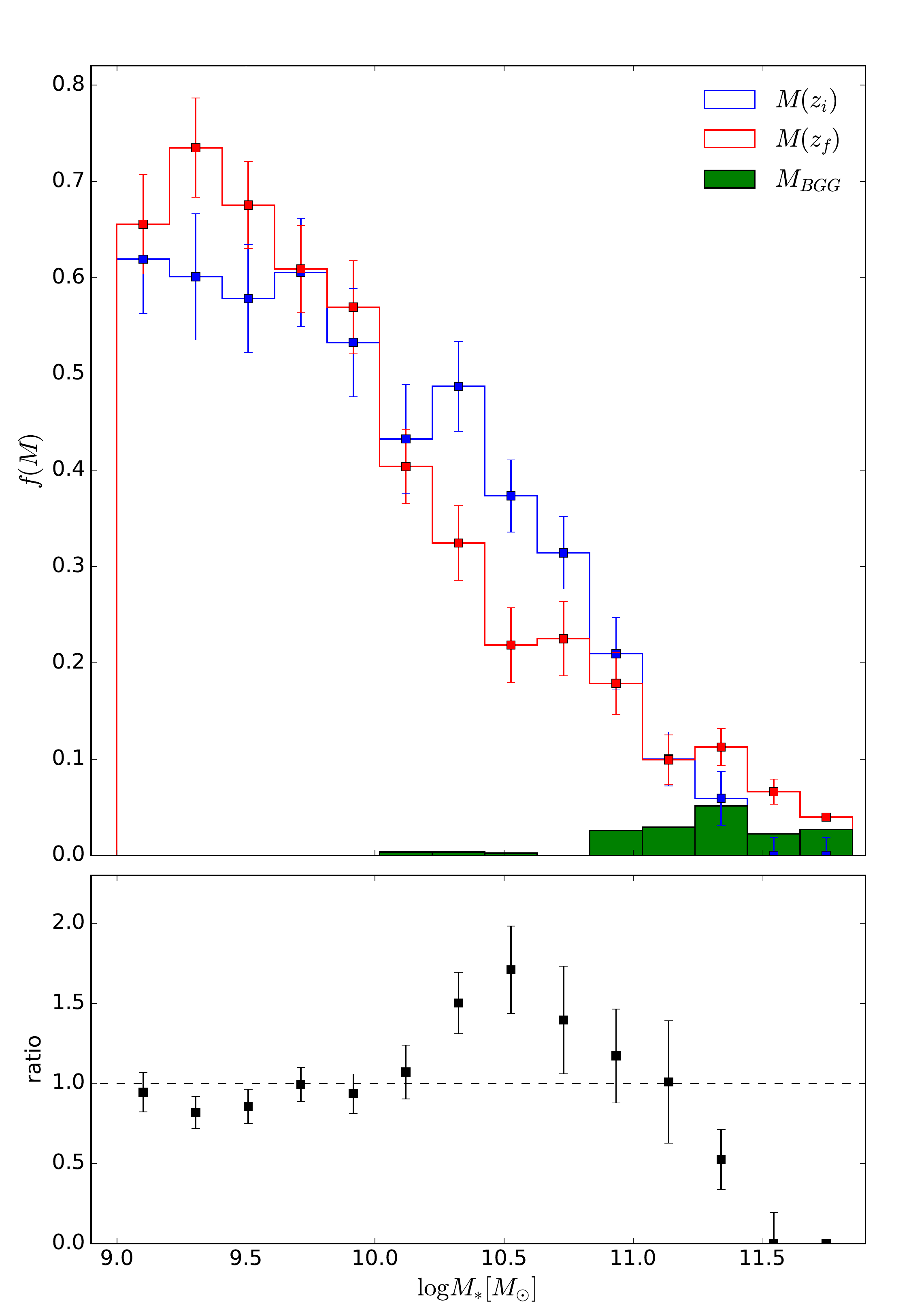} 
\caption{\small Evolution of the differential galaxy stellar mass (NIR luminosity) distribution function driven by the gravitational collapse of isolated small groups. \emph{Top:} Probability histograms for $\zi=3$ (blue solid line) and $\zf\simeq 0$ (red solid line) inferred from the stacking of our 48 synthetic galaxy groups. The area under both histograms is the same. The green solid histogram shows the stellar MF of BGGs. \emph{Bottom:} Ratio between the $\zi=3$ and $\zf\simeq 0$ probability histograms. Error bars depict $1\,\sigma$ uncertainties from bootstrap.}\label{fig_12}
\end{figure}

It is also instructive to examine the impact of the dynamics of group
formation on the long-term evolution of the stellar mass function, or
equivalently, the NIR LF of member galaxies. We do this in
Fig.~\ref{fig_12} where, in order to compensate for the small number
of objects in each individual simulated group, we have co-added
galaxies of similar stellar mass in our set of 48 collapsing groups to
create a single PDF. This plot reveals that towards the end of the
simulations, the composite luminous MF of galaxies develops a
statistically significant hump at the high-mass end driven by the
formation of first-ranked galaxies (most having $\ms > 10^{11}\msun$,
as shown by Fig.~\ref{fig_6}), which is accompanied by an also
statistically significant dearth in the number of objects with
intermediate mass. This dip in the moderate-mass regime
($M\lesssim\mc$) is reminiscent of the deficit of intermediate
luminosity galaxies found in the class of small and very compact HCGs
revealed in deep photometric surveys \citep*[e.g.][]{HCZ98}, and has
also been observed in the $B$- and $R$-band LFs of X-ray dim GEMS
(Group Evolution Multiwavelength Study) groups reported by
\citet{Mil04}, as well as in a very recent study of the study of the
shape of the LF using data obtained by the fossil group origins (FOGO)
project \citep{Zar15}. As shown by \citeauthor{Mil04}, the composite
LF of the low-$L_X$ GEMS groups shows a bimodal shape with a
bright-end hunch, a wide central depression of $\sim 2$--3 magnitudes,
and an upturn at the faint end, whereas that of the X-ray bright
counterparts ($L_X > 10^{41.7}$ erg~s$^{-1}$) can be fitted with a
single \citeauthor{Sch76} function over the entire range of the
data. Both because of the location of the dip, which starts near the
position of the 'knee' of the original distribution, and because of
its breadth, $\sim 2.5$ mag, and strength, the maximum deficit we
measure is close to $40\%$, we conclude that our simulated luminous MF
of galaxies at $\zf\sim 0$ is analogous to the bimodal LFs found in
the small galaxy systems of the local universe suspected of undergoing
rapid dynamical evolution.

\begin{figure*}
\begin{minipage}{170mm}
\centering
\includegraphics[width=0.8\linewidth,angle=0]{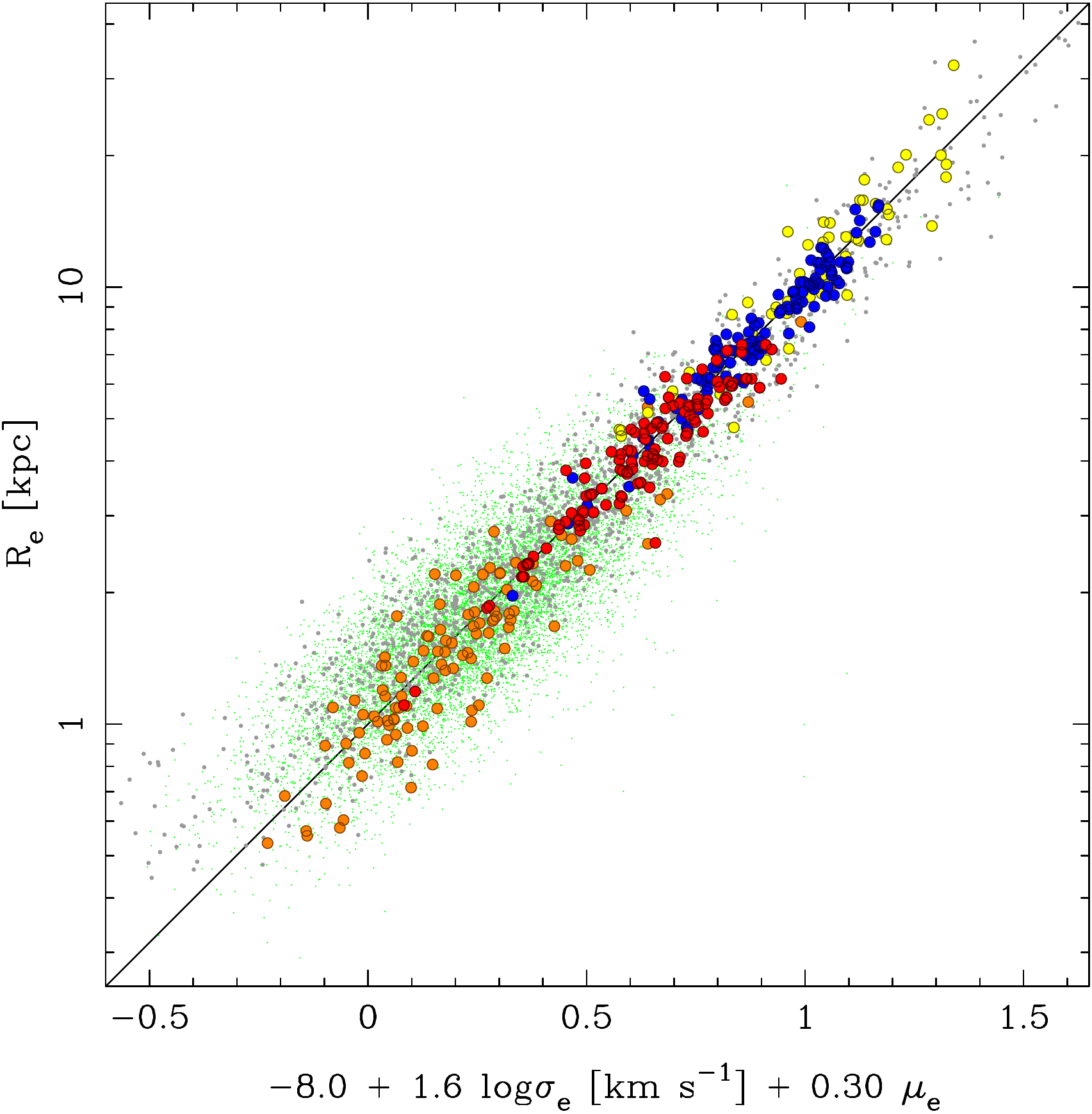}
\caption{Edge-on view of the FP of massive ETGs. The solid black line represents the best-fit to the mass FP defined by our mock first-ranked group galaxies (big dark-blue circles) in global parameter space $[\log\Reff,\log\sige,\SBe ]$. The distribution of our mock second-ranked galaxies is also shown (big red circles) together with four public data sets of nearby ETGs: a homogeneous set of 85 BCGs extracted from clusters (big yellow circles), the ETGs in the Norma and Coma clusters (big orange circles), the 6dFGSv catalogue (green dots), and a volume-limited sample of ETGs based on SDSS-UKIDSS observations (small grey circles). All subsets, simulated and real, show a strikingly similar tilt and normalization. Note that we plot three independent points for each mock galaxy corresponding to the estimates from the projections onto the three Cartesian planes.}\label{fig_13}
\end{minipage}
\end{figure*}

\section{Joint distribution of basic global properties: mock vs.\ real data}\label{scalings}

In this section we provide additional evidence favourable to the
hierarchical dry merging formation scenario for first-ranked galaxies
by comparing the joint distribution in 3D space of the most
fundamental properties of mock central remnants with real data.

\subsection{Selection of global parameters and control samples}\label{params&catalogs}

For each galaxy we considered estimators of the half-stellar mass
(-NIR-light) or effective radius, $\Reff$, the mean stellar velocity
dispersion within $\Reff$, $\sige$, and the mean stellar mass surface
density within $\Reff$, $\SBe \equiv -2.5\log[I_{\rm e}] =
-2.5\log[(\ms/2)/(\pi\Reff^2)]$ (for which we use the same symbols
usually adopted to represent surface brightness; see
Sec.~\ref{optical}). We replace mass by surface density not just
because the latter is a distance-independent quantity usually
preferred by observers, but also because it makes the FP independent
of scale, as its three dimensions are proportional to $M^{1/3}$. Note
also that the exact definitions of all these parameters are arbitrary
as long as they can be directly related to the values of the basic
properties of size, internal velocity and mass involved in the virial
theorem.

We also considered four control samples of nearby ETGs built using
different selection criteria. On one hand, we have chosen the
catalogue of ETGs in clusters Norma and Coma by \citet{Mut14}. These
authors report 2MASS $K$-band data derived from ESO New Technology
Telescope Son of ISAAC images on effective radii and surface
brightnesses, as well as velocity dispersions from Anglo-Australian
Telescope 2dF spectroscopy for 31 early-type galaxies in the Norma
cluster. For the Coma cluster, they use 2MASS images and SDSS velocity
dispersion measurements for 121 objects. Since the reported velocity
dispersions are corrected to the standardized physical aperture size
of $1.19$\;h$^{-1}$ kpc, we transform them to $1\Reff$ following
\citet{vdL07} to ensure consistency with the values measured from our
simulations. Total magnitudes are determined from surface brightness
profile fitting. A second data source is the 6dFGS Fundamental Plane
(6dFGSv) catalogue of nearly 9000 early-type galaxies in the local ($z
< 0.055$) universe \citep{Cam14}. In this data set, velocity
dispersions are derived from 6dF $V$-band spectral data, whereas the
photometric FP parameters are inferred from seeing-convolved
two-dimensional surface brightness fitting of the 2MASS images. The
published catalogue includes only spectral measurements with $\sigma >
100$\ \kms\ and lists velocity dispersions taken within $6.7$ arcsec
diameter fibre aperture, that we also correct to $1\Reff$. Another
data set of similar characteristics is the sample of 1430 ETGs defined
in \citet{LaB08}. These are galaxies extracted from a complete
volume-limited catalogue of the SDSS-DR5 database in the redshift
range $0.05$--$0.095$ which have ETG-like SDSS spectroscopic and
photometric parameters, reliable central velocity dispersions between
$70$ and $420$\ \kms, and available $K$-band photometry from the
second data release of the UKIRT Infrared Deep Sky Survey
(UKIDSS). Finally, we have selected the homogeneous set of 85 BCGs
with $z<0.1$ extracted from the C4 cluster catalogue with Sloan
$r$-band photometry reviewed by \citet{Liu08}. These authors list
velocity dispersions taken within a 3 arcsec fibre aperture and total
fluxes measured in terms of the $r$-band absolute magnitude at 25 mag
arcsec$^{-2}$. As in the previous cases, we correct $\sigma$ for
aperture size, while their magnitudes are converted into $K$-band
values by using the colour equation
\begin{equation}\label{SloantoK}
K=r-2.4(g-r)-1.15\;.
\end{equation} 
To derive eq.~(\ref{SloantoK}), we have first determined by
best-fitting to the multiband Coma cluster photometry \citep{Eis07} a
linear transformation between $(B-V)$ and $(K-R)$ colours, then used
the relations by \citet{FSI95} to transform from $(B-V)$ to $(g-r)$
and, finally, applied the conversion from $R$ to Sloan $r$ magnitudes
given by \citep{Win91}. The $K$ magnitudes resulting from this
equation, which adds an extra scatter of about 0.1 magnitudes, are
fully consistent with the stellar masses determined by the
prescription of \citet{Bel03} from $r$-band luminosities and $(g-r)$
colours. We do not attempt, however, to correct for differences in the
derivation of the total flux (i.e.\ model fitting versus isophotal
magnitudes).

\begin{table}
\scriptsize
\centering
\caption{Scatter about the Fundamental Plane defined by our mock BGGs}
\label{FPscatter}

\begin{tabular}{lccc}
\hline \hline
Data Set & $\log\Reff$ & $\log\sige$ & $\SBe$ \\
\hline 
Simulated BGGs & 0.04 & 0.03 & 0.14\\
2nd.-ranked mock galaxies & 0.05 & 0.03 & 0.18\\
Observed BCGs & 0.08 & 0.05 & 0.25\\
Cluster spheroids & 0.10 & 0.06 & 0.32\\
General ETG population & 0.14 & 0.08 & 0.45\\ \hline
\end{tabular}
\end{table}

\subsection{The mass Fundamental Plane}\label{massFP}

Fig.~\ref{fig_13} shows an edge-on view of the mass FP defined by our
mock first-ranked group galaxies (big dark-blue circles), fitted using
orthogonal least squares, in the global parameter space
$[\log\Reff,\log\sige,\SBe ]$:
\begin{equation}\label{orthoFP}
\begin{split}
\log\Reff & =(1.60\pm 0.05)\log\sige+\\ &+(0.75\pm 0.02)\SBe/2.5-8.0\pm 0.2\;,
\end{split}
\end{equation}
where the quoted errors are median absolute deviations from $4,000$
bootstraps. We have also plotted in this figure the $K$-band data
points corresponding to the different classes of spheroidal systems
included in the control data sets, as well as as to our second-ranked
mock galaxies. The FP inferred from our simulations matches
exceptionally well not just the mean tilt, but also the intercept of
the relatively narrow flat 3-space surface delineated by the total set
of observational data. The excellent visual agreement between data and
simulations is quantitatively supported by the full consistency,
within the statistical uncertainties, shown between
equation~(\ref{orthoFP}) and the functional forms quoted in the three
papers cited above dealing with control samples that use $K$-band
photometry. In the case of the \citet{Liu08} BCGs' sample, however, it
is not possible to compare the fits directly. Finding the best-fitting
model of a set of measurements is a complex problem that requires
careful modelling of sample selection effects and a good determination
of observational errors \citep[see, e.g.][]{Tor11}. Since both are
conditions difficult to fulfill with data coming from external
sources, we have chosen to assess the level of agreement between the
our FP and this fourth data set using only a qualitative
approach. Thus, its is easy to convince ourselves that
Fig.~\ref{fig_13} also provides a reasonable edge-on view of
\citeauthor{Liu08}'s observations based on the extreme narrowness
exhibited by the BCGs point cloud perpendicularly to the direction
delineated by our FP, which runs essentially through the data
midsection with no noticeable signs of an offset.

Having established that equation~(\ref{orthoFP}) provides a reasonably
good description of all the observed samples, we have grounds on which
to estimate the r.m.s.\ deviations arising from the comparison of the
response along each dimension, calculated from
equation~(\ref{orthoFP}) using the other two dimensions as predictor
variables, with the actual measurements. As Table~\ref{FPscatter}
reveals, the scatter around the FP is connected to the ETG class,
increasing as the typical stellar mass of the representatives of a
class decreases. Our simulated massive central galaxies, as well as
their second-ranked companions, have small residuals of typical size
somewhat lower than those shown by the BCGs' data. Overall, we find
that these extreme galaxies are distributed more tightly around the FP
than the bulk of spheroids in clusters, which in turn show a smaller
scatter than that of the general ETG population. This is but another
manifestation of the important regularity in the formation process of
massive hot spheroids. The low dispersion of the measurements from our
highest-ranked galaxy models demonstrate, in good harmony with other
recent investigations of stochastic collisionless merging
\citep{TDY13}, that this mechanism is capable of producing tight
relations between fundamental parameters. It appears then that a
realistic dissipationless merger hierarchy can naturally provide the
conditions required to reproduce the small intrinsic scatter shown by
the observed stellar-mass scaling laws \citep{Nip09}. In view of the
fact that such scalings not only quantify the behaviour and nature of
first-ranked group galaxies, but are a primary tool for validating the
formation path proposed for this class of objects, we have decided to
pay to this particular issue the attention it deserves. For this
reason, we do not to expand the present discussion any further and
refer the interested reader to the more thorough analysis of the
stellar-mass FP defined by our mock BGGs, its 2D projections in $RVM$
space, and their compatibility with the corresponding scalings
observed in the local volume from different kinds of giant ETGs, that
we present in a companion paper \citep{PS16}.

\section{Main results and conclusions}\label{conclusions}

We have run a large suite of controlled multiple dry merger
simulations of galaxies of various masses and morphologies in
previrialized groups with twenty-five initial members. Our galaxy
groups are evolved from an initial epoch, in which they behave
essentially as a linearly expanding homogeneous perturbation, until
the end of their first fully non-linear gravitational collapse, when
the surviving galaxies adopt a compact configuration. These
experiments aim to provide a large sample of cosmologically realistic
hierarchical merger histories that allow for a fair assessment of the
role played by gravity in the build-up of the most massive remnants of
densely populated environments.

Under the hierarchical structure formation paradigm it is natural to
expect that the observable properties of BGGs bear some relationship
to the properties of the overdensity on which they have formed. We
have found that in our previrialized groups both the magnitude of the
first-ranked remnant and its \emph{degree of dominance}, as measured
by the magnitude difference between this and the next ranked galaxy,
are positively correlated with the global stellar mass fraction of the
parent halo, and hence with the initial total mass in bound subhaloes,
the most determining property that controls the assembly histories in
our simulations. The second of these relationships appears to be
independent of the well-established covariance of the gap size with
richness \citep[e.g.][]{LS06}, which suggests that these two later
parameters are also dynamically connected. As pointed out by
\citet{Hea13}, Bayes' Theorem implies that if BGG dominance were
purely statistical, i.e.\ resulting from $N$ random draws from a given
LF function, the magnitude gap between the two brightest galaxies
would contain no information about mass that would not already be
provided by the knowledge of richness. Actually, if one excludes the
effect of sparse sampling of the LF, which biases high such dominance
-- and may partly explain the observed tendency of large-gap systems
to have fewer members than their small-gap counterparts --, the
results of the present work point to the fact that richness would act
only as a secondary player in establishing the breadth of
$\Delta\magn_{12}$, which would be driven mainly by the initial mass
distribution among the group members. As previously determined in the
numerical modelling of massive clusters \citep[e.g.][]{Gao04}, we have
found evidence that the dominance of the first-ranked members in
dynamically evolved galaxy systems is essentially built up during the
gravitational collapse that precedes their virialization. Our
simulations show that the size and mass growth of a large fraction of
central galaxies occurs at the expense, in a literal sense, of their
lesser companions: many of our mock BGGs have been found to arise from
the accretion into the largest precursor galaxy of mid-sized
progenitors, the number of which is well correlated with the final
luminosity achieved by the BGG. This implies that the brightening
process of central remnants is for the most part special and therefore
explains why their luminosities are generally inconsistent with a
statistical extreme value population \citep{She14}. Besides, we have
observed that the preferential destruction of intermediate-mass
objects produces a characteristic dip in the luminous MF of galaxies
which bears many similarities with the central depression observed in
the optical/NIR galaxy LF of X-ray dim groups by \citet{Mil04} and
other dynamically young systems. The connection between the presence
of large gaps in the MF/LF of galaxy systems and the growth via
dissipationless, dynamical friction-driven merging of their central
galaxies has been recently confirmed by observational studies
\citep{Zar15}. In fact, the idea that the most massive galaxies in
locally dense environments form at the expense of their moderately
bright neighbours is already present in early theoretical treatments
of merging, like the one developed by Cavaliere and co-workers as a
complement of the canonical hierarchical clustering in galaxy
aggregations \citep[e.g.][]{CM97,Men02}. Such a model shows that
multiple binary collisions of satellite galaxies in low mass groups
can effectively deplete the number of intermediate-mass objects
(i.e.\ those having internal stellar velocities $V\sim
50$--200~\kms\ at the scales of interest), while barely affecting the
abundance of the smallest satellites. This behaviour stems from the
fact that the average merger rate, which is proportional to
$1+(V/V_{\rm{rel}})^2$, with $V_{\rm{rel}}$ the relative velocity of
galaxies within the parent halo, becomes gradually negligible for the
galaxies having the lowest $V$ as the secular evolution of clustering
makes the entire galaxy system progressively more massive. In
addition, disrupted peripheral baryons coming from the destroyed
galaxies may contribute to the generation of diffuse IGL which, in the
case that includes new stars, can help to explain the presence of
intragroup material bluer than the mean colour of the group galaxies,
like in HCG79 \citep{DRMO05}.

The singular formation route of first-ranked objects also leaves its
imprint on their internal configuration. Although this is an issue
that we expect to address in more detail in future analyses, we have
found some results regarding their light distribution that are worth
commenting. Thus, our simulations show that, in contrast to the
findings of similar former experiments \citep[see, e.g.][]{TDY13}, the
stellar structure of central galaxies assembled during the
virialization of their parent haloes -- quantified by the values of
the S\'ersic index $n$ of their integrated light profiles -- bears no
obvious relation to the morphologies and number of mergers/input
progenitors, more numerous on average than for lesser group
members. These results, and other checks done with our data, indicate
that the final shapes of the BGGs, in most cases similar to those of
classical ellipticals \citep*{GZZ05,Ber07}, are not tied to the nature
of the merging objects, nor to the characteristics of the parent
galaxy system. This evidence suggests that the self-similarity in the
profiles of our central galaxies comes primarily from the relatively
large amount of orbital energy that our simulated mergers are expected
to turn into random stellar motion during the formation of the host
groups.

Finally, we have measured the most basic global properties of our mock
central group remnants and compared their joint distribution in the
global parameter space $[\log\Reff,\log\sige,\SBe ]$ with real data
from the local volume. We find that our BGGs define a thin mass FP
relation which matches perfectly well both the slope and normalization
of the distributions of different kinds of massive ETGs inferred at
NIR wavelengths, thus providing a unifying relation uniting all kinds
of giant ellipticals, from ordinary objects to the brightest members
of groups and clusters. This successful reproduction of the observed
tilted FP relation has not required the presence of gas in the
progenitor galaxies. This is in sharp contrast with former
dissipationless merger experiments of discs that, which with the
exception of the early results by \citet{AV05} and the much more
recent by \citet{TDY15}, have systematically produced remnants that
occupy a FP similar to that expected from the standard relation for
virialized homologous systems: $R\propto\sigma^2I^{-1}$ \citetext{see
  also \citealt{PS16}}.

Our dissipationless simulations of previrialized groups indicate that
the formation of BGGs can be driven by the extensive (and intensive)
merging accompanying the gravitational collapse of the parent
structure. Therefore, the present experiments agree well with
observational findings suggesting that massive ellipticals in the
local universe have mostly grown from the inside out through dry
mergers in the last 10 Gyr \citep[see, e.g.,][and references
  therein]{vDok10}. They are also conceptually consistent with the
"two-phase" scenario predicted in various studies on the formation
histories and evolution of these objects in a cosmological context
\citep[e.g.][]{DeLB07,Ose10}, in which most stars that end up in large
galaxies are formed early within smaller structures, whereas the
galaxies themselves are assembled at a later epoch \citep[but see,
  however,][]{N-G13}. Specifically, \citeauthor{Ose10} propose that
first-ranked galaxies grow considerably in mass and radius during an
extended phase starting at $z \lesssim 3$ of "ex situ" accretion and
merging with smaller stellar systems (satellites) created in earlier
times outside the virial radius of the forming central object. Their
conclusions are based on the outcome of fully cosmological
high-resolution re-simulations of 39 individual galaxies with
present-day virial halo masses ranging from $\sim 7\times
10^{11}$--$3\times 10^{13}\,h^{-1}\msun$. In these experiments
galaxies accrete between about $60\%$--$80\%$ of their present-day
stellar mass (the upper limit corresponding to galaxies with
$\ms\gtrsim 10^{11}\,h^{-1}\msun$), with most of the assembly taking
place towards lower redshift ($z\lesssim 1$). For the most massive
systems this mass growth is not accompanied by significant star
formation. On the other hand, the majority of their "in situ" created
stars are formed quite early \citetext{$z>2$; according to
  \citet{DeLB07} up to $80\%$ of the stars are formed at $z\sim 3$},
as observations suggest, fuelled mostly from cold gas flows. This
scenario would also explain the seemingly anti-hierarchical behaviour
that arises from the fact that the oldest stellar populations reside
in most massive ETGs \citep[e.g.][]{DeL06}.

The results of the present controlled multiple merger simulations,
together with those from former hydrodynamical experiments, suggest
that dry hierarchical -- increasingly minor \citep*{HNO13,Moo14} --
merging and dissipational merging should not be regarded as mutually
exclusive pathways for the assembly of ETGs, but actually as
supplementary scenarios. Our view is that both are essential to
develop a complete physical understanding of the formation of the
whole population of elliptical galaxies and the origin of the tight
scaling relations that connect their global properties. The
predominance of either formation route would be determined mostly by
the local environment. In groups and proto-cluster regions, frequent
collisions and a potential shortage of new cold gas would work
together to assemble massive, slow-rotating red galaxies from a
sequence of multiple and sometimes overlapping dry mergers of systems'
members -- a process that certainly turns out to be particularly
intense for the central remnants of groups and clusters, even compared
to other similarly massive ETGs. On the other hand, fast-rotating,
bluer ETGs of intermediate luminosity and discy isophotes would form
via dissipational (essentially pairwise) merging, accompanied by a
compact central starburst \citep{HCH08}, mostly from gas-rich LTGs
that abound in low-density environments. The scenario we are
suggesting is, to some extent, reminiscent of the \emph{merging
  continuum} idea that the importance of dissipation on the merger
process must decline with increasing galaxy mass, proposed by
\citet*{BBF92} \citep[see also][]{Kor89} as a way to reconcile the CDM
predictions on structure formation with data from the most luminous
ellipticals in groups. Whatever the case, it is no less striking that
such varied assembly routes for ETGs, as regards as to the nature of
the progenitors and their expected impact on the evolution of the
dark-matter-to-stellar ratio of the remnants, end up leading to a
unique tight scaling of their most basic properties.

\section*{Acknowledgments}

The authors wish to thank the referee, Dan Taranu, for his thorough
revision of the manuscript and insightful comments and suggestions,
which have helped to improve both the justification of the initial
conditions of our simulations and the presentation of the
results. This work was supported by the Program for Promotion of
High-Level Scientific and Technical Research of Spain under contracts
AYA2010-15169 and AYA2013-40609-P. The simulations of the present work
have been performed with the Computer Service Infrastructure of the
Spanish Instituto de Astrof\'\i sica de Andaluc\'\i a (IAA).

This research is part of the IDILICO (Investigation of the DIffuse
LIght COmponent in compact groups of galaxies) project, an
international collaboration that runs simulations of the formation of
groups of galaxies using high-performance computing resources (see our
website at
\texttt{http://www.am.ub.edu/extragalactic/idilico/}). Initially born
to shed light on the origin of the unusually large fractions of
diffuse intragroup light found in some compact groups of galaxies,
IDILICO has now among its goals the creation of a comprehensive and
unbiased sample of forming galaxy associations suited for
investigating the aspects of galaxy evolution governed by
gravitational dynamics, especially those driven by the multiple
interactions and mergers that galaxies undergo in dense
environments. Subsets of our numerical experiments are available upon
request.








\appendix

\section{Two-body heating in our galaxies}\label{num_resol}

In this Appendix, we turn to an inspection of the effects of numerical
heating in our mock galaxies. This is done by running in isolation two of
our (elliptical and disc) galaxy models at different resolutions.
Specifically, we have evolved one elliptical model and one disc model
with about $315,000$ stellar particles each, aimed to represent
galaxies with a total mass of $1\;\mc$ in our simulations, as well as
versions of both kinds of galaxies with a factor ten drop in
resolution, in this case intended to portray Magellanic-cloud-sized
objects with a total mass of $0.1\;\mc$. A fifth simulation that
reproduces again the same elliptical galaxy model, but using this time
only $\sim 15,000$ stellar particles, has been run to study the
stability of the most extremal low-resolution objects included in our
simulations. The settings of the different components in terms of
mass-resolution, softening of the gravitational potential and timestep
are the same used for the group simulations. Numerical stability is
investigated by means of a simple test consisting in following for a
total of six simulation time units (i.e., nearly 12 Gyr) the evolution
of both the characteristic size of the luminous component of the
galaxies, $\Reff$, directly measured from particle counts projected
onto the main symmetry plane of the galaxies, and the average
l.o.s.\ velocity dispersion within this radius, $\sige$.

\begin{figure}
\centering
\includegraphics[width=\linewidth,angle=0]{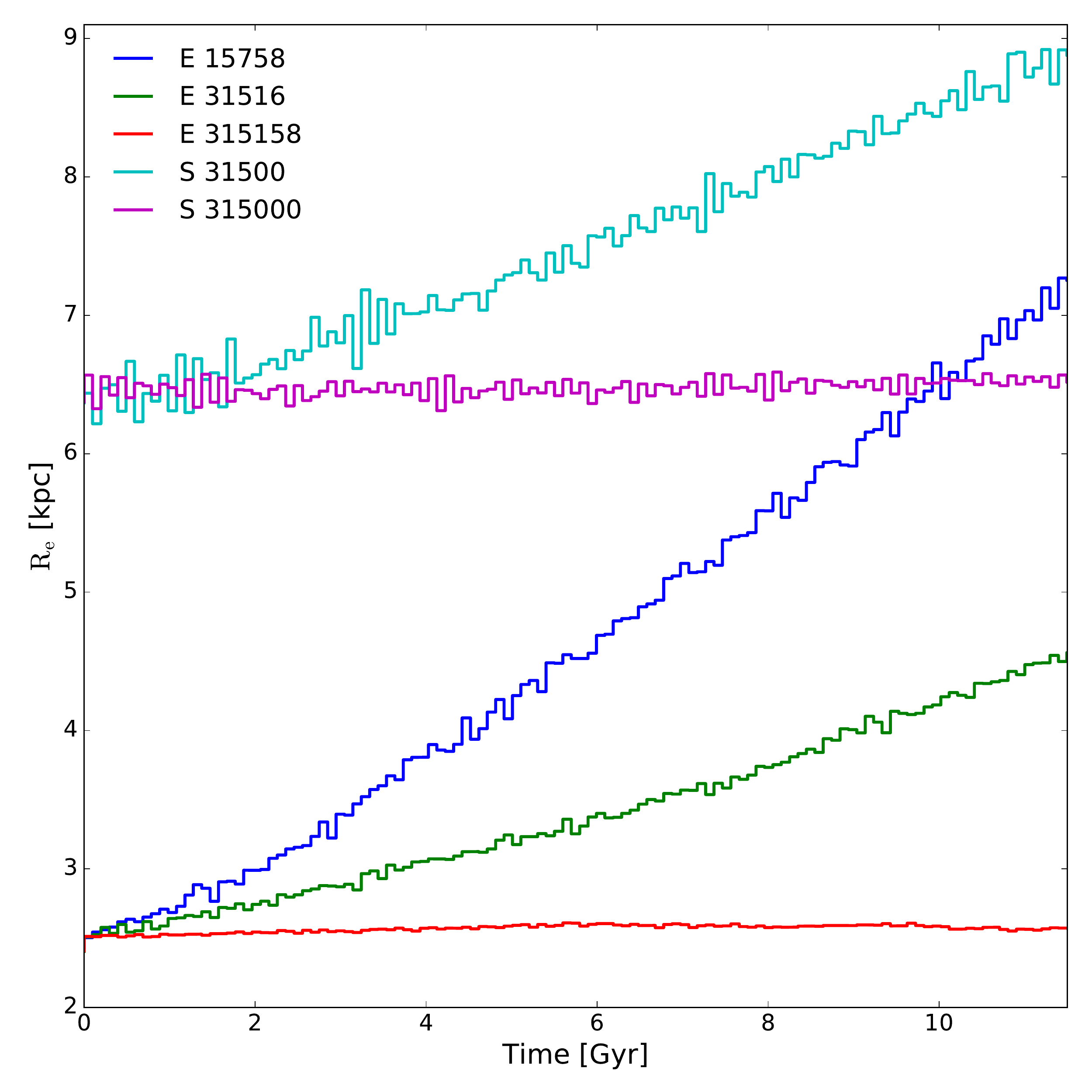} 
\caption{\small Secular evolution of the projected characteristic size
  of the luminous component, $\Reff$, of one of our elliptical (E) and
  one of our spiral (S) galaxy models built using different
  resolutions. The labels of the curves inform about the morphology of
  the galaxy and the number of stellar particles it
  contains.}\label{fig_14}
\end{figure}

As shown in Fig.~\ref{fig_14}, the secular variation of parameter
$\Reff$ is basically linear throughout most of the whole testing
process and, as expected, strongly resolution dependent. We find that
the size of our highest resolution models changes very little, with
average fractional growths at the $0.5\%$ level per Gyr for the
$R^{1/4}$ spheroid, while the $\Reff$ of the exponential disc remains
essentially constant even after 12 Gyr, with oscillations between
consecutive values that in no case exceed $2\%$. On the other hand,
the size of our moderately resolved galaxies made with ten times less
particles experiences only a $3\%$ variation in the first 2 Gyr for
the disc model, followed by an steeper increase at an average rate of
about $3\%$ per Gyr. This last figure approximately doubles for its
elliptical counterpart, leading to a cumulative deviation of about
$30\%$ ($15\%$ for the disc model) after 6 Gyr, the typical time after
which the merger rate in our groups becomes significant. These
values can be compared with the intrinsic dispersion in the
(logarithm) of the distribution of early-type galaxy sizes at fixed
absolute magnitude, $\Delta \log \Reff\approx 0.125$ \citep{Ber03a},
which is near $40\%$ for the typical value of $\Reff$.

As regards the variations in $\sige$, they behave similarly to those
of $\Reff$, but with elliptical and discs exchanging roles regarding
the amplitude of the variations. Thus, this parameter remains
essentially constant for our highest-resolution elliptical model after
12 Gyr, while for the disc model $\sige$ increases at a modest rate of
around $1.5\%$ per Gyr. For the moderately resolved models we record
increases of less than one per cent for the elliptical model and up to
$2.5\%$ per Gyr for the disc one.

Finally, in the case of the centrally concentrated $15$k-star particle
model, we find that numerical heating becomes a more serious
issue. The size parameter of such a low-resolution spheroid
deteriorates at a rate about twice as fast on average than that of
their $30$k-star counterparts, approaching a 70 per cent variation in
6 Gyr. In contrast, their central velocity dispersion is much more
stable, registering total increments that barely exceed $7\%$ in the
same period of time\footnote{Also note that this simple test provides
  an upper limit to the effects of two-body heating, as in our group
  simulations the concentration of the galaxies increases with
  decreasing mass.}. In either case, we recall that these last numbers
are not a real concern since, as has been already stressed on the main
text, galaxies that small play a merely testimonial role in the
formation process of the BGGs investigated in the present work.

\section{Evolution of halo properties with cosmic time}\label{halo_evolution}

In this Appendix, we describe the model adopted for the growth of our
galaxy dark haloes and comment on its implications on the time
dependence of the fundamental shape parameters of this component.

A rough estimate of the mass aggregation rate to which our mock
galaxies may be subject during the evolution of their parent groups
can be inferred from the physical model by \citeauthor*{SMS05}
\citetext{\citeyear{SMS05}; see also \citealt*{Man03}}. This model is
based on the premise that the universal spherically averaged density
profile of relaxed dark matter haloes does not depend on the halo
aggregation history. Such a property stems from the fact that the
parameters used to characterize the dark haloes of galaxies, namely
the scale radius $\rs$ and its directly related mass $\Ms$ within it,
are determined at any redshift by essentially the current total mass
and energy of the haloes, as well as by their respective instantaneous
accretion rates. This means that relaxed haloes emerging from major
mergers, which lead to the structural rearrangement of their
progenitors, should show radial profiles indistinguishable from those
of haloes with identical global properties (in particular, identical
virial mass) and boundary conditions, but having endured instead a
more gentle growth. As \citet {SMS05} show, this is an approach fully
consistent with the results of high-resolution cosmological $N$-body
simulations.

The preservation of the inner structure of dark haloes predicted by
this inside-out growth scheme \citep[see also][]{LP03,B-K09}, provides
the means to estimate the typical fractional mass increase endured 
by a galactic halo along a given period of time. To do this in a
manner independent of the exact form of the dark halo profile, we
combine the assumed invariance of $\rs$ between the initial and final
redshifts of our simulations, with the expression for the halo virial
radius in Eq.~(\ref{Rvir}), the definition of the concentration
parameter and the mass concentration relation in
Eq.~(\ref{M-c}). Then, it is straightforward to show that for haloes
of any mass
\begin{equation}\label{massincr}
\frac{\Mvir(\zi)}{\Mvir(\zf)}=\left[\frac{\Dvir(\zi)}{\Dvir(\zf)}\right]^{1/(1-3\gamma)}\;,
\end{equation}
where $\gamma =-0.094$ is the
power-law index of the median $M$--$c$ relation adopted (Eq.~(\ref{M-c})).

According to Equation~(\ref{massincr}), the fractional increase of the
mass of a relaxed galactic halo between $\zi$ and $\zf$ depends only
on the cosmology under consideration (through the ratio of $\Dvir$'s
and $\gamma$)\footnote{This halo growth model preserves the
  ratio $\Mvir/\mc$ at all $z$ and, hence, the halo MF in reduced mass
  units. This is equivalent to state that there is no differential
  evolution in the MF of haloes.}. When applied to our
simulations, Equation~(\ref{massincr}) implies that the $\zi=3$
progenitors of $\zf=0$ galaxies are objects with virial masses
typically a factor $0.578$ lower. The corresponding reduction in the
virial radius can be calculated from Equation~\ref{Rvir}.

\section{Density evaluation}\label{deneval}

All density maps presented in this paper have been obtained
calculating the mass surface density using a smoothed particle
hydrodynamics kernel on the projected particle distributions, in the
form \citep[cf.][]{DA12}
\begin{equation}\label{densest}
\rho(\mathbf{x})\approx\hat{\rho}\equiv\sum_i m_iK((\mathbf{x}-\mathbf{x}_i);h_i)\;,
\end{equation}
where $\hat{\rho}$ is the density estimate at the coordinate
$\mathbf{x}$ of the particle/pixel, $K(\Delta\mathbf{x};h)$ the
smoothing kernel and $h$ the smoothing scale. Specifically, we use the
\citeauthor{Wen95} isotropic C2 kernel with compact support
\begin{equation}\label{C2kernel}
\begin{split}
& K(\Delta\mathbf{x};h)= \\ & \ \ \left\{ \begin{array}{ll}
               \frac{7}{64\pi h^2}\left( \frac{|\Delta\mathbf{x}|}{h}-2 \right)^4\left( 2\frac{|\Delta\mathbf{x}|}{h}+1 \right),
               & \mbox{if $\frac{|\Delta\mathbf{x}|}{h}\leq 2$,} \\[0.1em] \\
               0, & \mbox{otherwise,}
               \end{array}
               \right.
\end{split}
\end{equation}
where for two dimensions
$|\Delta\mathbf{x}|=\sqrt{(x-x_i)^2+(y-y_i)^2}$). We have fixed $h$ to
ten times the Plummer-equivalent softening length of star particles
($\epsilon=30$~pc), deliberately ignoring the recommended adaption of
this parameter such that for $\nu$ dimensions
$h_i^{\nu}\hat{\rho}\approx\mbox{constant}$ \citep{Li06}. We have done
so motivated by the large number of bodies in our simulations, which
is sufficient to warrant a low noise level, as well as by our desire
to visually enhance the contrast of the lowest density substructures
of intergalactic light that are created as a result of galaxy
collisions (see Fig.~\ref{fig_3}). Adaptive smoothing will be used in
future work when we study in detail the structure and kinematics of
massive merger remnants.

For calculations done using single particle positions it becomes
necessary to speed up density evaluation. In such cases, we first
create a partition of the two-dimensional space by recursively
subdividing it into four quadrants containing at most a prefixed
number $N_ {\rm cell}$ of bodies. In this manner, for each particle we
only need to locate the adjacent cells containing particles that may
enter in the sum~(\ref{densest}). This dramatically reduces the number
of inter-particle separations that must be computed.



\bsp	
\label{lastpage}
\end{document}